%% file: main.tex
\documentclass[lettersize,journal]{IEEEtran}
\usepackage{amsmath,amsfonts}
\usepackage{algorithmic}
\usepackage{algorithm}
\usepackage{array}
\usepackage[caption=false,font=normalsize,labelfont=sf,textfont=sf]{subfig}
\usepackage{textcomp}
\usepackage{stfloats}
\usepackage{url}
\usepackage{verbatim}
\usepackage{graphicx}
\usepackage{cite}
\usepackage{comment}

\PassOptionsToPackage{svgnames}{xcolor}
\usepackage{xcolor}
\usepackage{tikz}

\usepackage{braket}

\usepackage{mfirstuc}

\usepackage{gensymb}
\usepackage{soul}

\usepackage{tgheros}

\definecolor{myColor}{RGB}{99, 99, 99}
\newcommand*\component[1]{\tikz[baseline=(char.base)]{
            \node[rectangle,draw=myColor, line width=0.5mm,rounded corners=0.7mm,text=white, fill=myColor, inner sep=0pt,minimum size=11pt](char) {\fontfamily{qhv}\selectfont{#1}}}}
\newcommand*\subcomponent[1]{\tikz[baseline=(char.base)]{
            \node[rectangle,draw=myColor, line width=0.5mm,rounded corners=0.7mm,text=white, fill=myColor, inner sep=0pt,minimum size=10pt](char) {\textsf{\footnotesize #1}}}}
\definecolor{emphColor}{RGB}{232,232,232}
\newcommand*\emphasize[1]{\tikz[baseline=(char.base)]{
            \node[shape=rectangle,fill=emphColor, draw=olive, text=black, inner sep= 2pt,minimum size=11pt,rounded corners=3pt] (char) {#1}}}

\definecolor{BoneColor}{RGB}{170,207,217}

\definecolor{BtwoColor}{RGB}{241,174,123}

\definecolor{BthreeColor}{RGB}{97,160,184}

\usepackage{graphicx}

\newcommand{\annotation}{\includegraphics[scale=0.25]{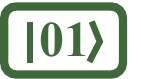}}%
\newcommand{\hub}{\includegraphics[scale=0.2]{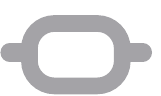}}%
\newcommand{\lightblueblock}{\includegraphics[scale=0.13]{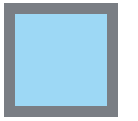}}%
\newcommand{\blueblock}{\includegraphics[scale=0.13]{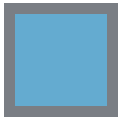}}%
\newcommand{\arrow}{\includegraphics[width=10pt, height=9pt]{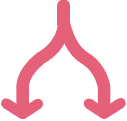}}%

\newcommand{\toolName}{\textit{QuantumEyes}}

\newcommand{\designName}{\textit{dandelion chart}}

\newcommand{\designURL}{\textcolor{blue}{\url{https://www.npmjs.com/package/dandelion_chart}}}
\newcommand{\systemURL}{\textcolor{blue}{\url{https://quantumeyes.github.io/}}}

\newcommand{\revise}[1]{\textcolor{black}{#1}}

\graphicspath{{figs/}{figures/}{pictures/}{images/}{./}} % where to search for the images

\usepackage{hyperref} %<--- Load after everything else
\begin{document}

\title{\toolName{}: Towards Better Interpretability of Quantum Circuits}

\author{Shaolun Ruan \orcid{0000-0002-6163-9786}, 
Qiang Guan \orcid{0000-0002-3804-8945}, 
Paul Griffin \orcid{0000-0002-1656-421X}, 
Ying Mao \orcid{0000-0002-4484-4892}, and 
Yong Wang \orcid{0000-0002-0092-0793}
\thanks{S. Ruan, Y. Wang, and P. Griffin are with Singapore Management University. E-mail: slruan.2021@phdcs.smu.edu.sg, \{yongwang, paulgriffin\}@smu.edu.sg}
\thanks{Q. Guan is with Kent State University. E-mail: qguan@kent.edu}
\thanks{Y. Mao is with Fordham University. E-mail: ymao41@fordham.edu}
\thanks{Y. Wang is the corresponding author.}
}

\markboth{Journal of \LaTeX\ Class Files,~Vol.~14, No.~8, August~2021}%
{Shell \MakeLowercase{\textit{et al.}}: A Sample Article Using IEEEtran.cls for IEEE Journals}


\maketitle

\begin{abstract}
%% Part 1: Emphasize the importance of quantum circuits and the importance of explaining quantum circuits
%Quantum computing has shown superior speedup over classical computing and it has motivated a large number of users to learn and apply quantum computing to various applications.
Quantum computing offers significant speedup compared to classical computing, which has led to a growing interest among users in learning and applying quantum computing across various applications. 
However, quantum circuits, which are fundamental for implementing quantum algorithms, can be challenging for users to understand due to their underlying logic, such as the temporal evolution of quantum states and the effect of quantum amplitudes on the probability of basis quantum states.
% e.g., how is the final probability generated and how are the quantum states evolving.
% Quantum computing has great potential to revolutionize computational statistics and data science considerably.
% Meanwhile, it has become increasingly popular in recent years due to its utilization in numerous emerging applications.
% Despite its increasing prevalence, it is still not trivial to learn and understand the internal logic hidden in quantum circuits, the most fundamental computational routine to realize the functionality of quantum algorithms.
% Thus, an approach that can assist users in better understanding quantum circuits is urgently needed considering the rapid expansion of novices and users in quantum computing.
%% Our solutions
To fill this research gap, we propose \toolName{}, an interactive visual analytics system to enhance the interpretability of quantum circuits through both global and local levels.
For the global-level analysis, we present three coupled visualizations to delineate the changes of quantum states and the underlying reasons:
a Probability Summary View to overview the probability evolution of quantum states;
% the quantum circuit's temporal evolution regarding the probability of each basis state;
a State Evolution View to enable an in-depth analysis of the influence of quantum gates on the quantum states;
a \revise{Gate Explanation View} to show the individual qubit states and facilitate a better understanding of the effect of quantum gates.
% by visualizing the individual qubit states.
For the local-level analysis, we design a novel geometrical visualization \designName{}
% which explicitly reveals 
to explicitly reveal
how the quantum amplitudes affect the probability of the quantum state.
% by several geometrically-correlated visual channels.
% and a Probability Explanation View with a novel geometrical visual design \designName{} to reveal how quantum amplitudes change the quantum state probability.
% To this end, we proposed a visualization system \toolName\ for visually enhancing the interpretability of quantum circuits.
% Specifically, \toolName\ provides a holistic picture of the evolution of quantum states throughout a quantum circuit via three coordinated views:
% a Probability Summary View overviews the quantum circuit's temporal evolution regarding the probability of each basis state,
% a State Evolution View allows the in-depth analysis of how each quantum gate affects the basis states for each step,
% a \revise{Gate Explanation View} assists users in better understanding the gate's effect from the perspective of the individual qubit.
% Moreover, \toolName\ further visually explains the quantum gates' behavior by Probability Explanation View with a novel geometrical design \designName.
% In particular, \designName\ explicitly reveals how amplitudes affect the probability of the quantum state by a set of geometrically-correlated visual channels.
% Evaluation
We thoroughly evaluated \toolName{}
as well as the novel \designName{} integrated into it 
through two case studies on different types of quantum algorithms and
in-depth expert interviews with 12 domain experts.
% We extensively evaluate \toolName\ and the embedded \designName\ through two case studies on different types of quantum programs and
% % well-designed 
% expert interviews with 12 domain experts.
The results demonstrate the effectiveness and usability of our approach in 
% achieving the interpretability improvement of quantum circuits.
enhancing the interpretability of quantum circuits.
\end{abstract}

\begin{IEEEkeywords}
Interpretability, data visualization, quantum circuits, quantum computing.
\end{IEEEkeywords}

\input{sections/1-introduction}
\input{sections/2-related_work}
\input{sections/3-background}
\input{sections/4-informing_the_design}

\input{sections/5-method}

\input{sections/6-case_study}
\input{sections/7-expert_interview}
\input{sections/8-discussion}

\input{sections/9-conclusion}

\section*{Acknowledgement}
% This research was supported by the Singapore Ministry of Education (MOE) Academic Research Fund (AcRF) Tier 1 grant (Grant number: 20-C220-SMU-011).
This research was supported by the Lee Kong Chian Fellowship awarded to Yong Wang by Singapore Management University.
We would like to thank Songheng Zhang and Cheng Chen for the proofreading and anonymous reviewers for their useful feedback.

% \section{References Section}
% You can use a bibliography generated by BibTeX as a .bbl file~\cite{liu2018deeptracker}.
%  BibTeX documentation can be easily obtained at:
%  http://mirror.ctan.org/biblio/bibtex/contrib/doc/
%  The IEEEtran BibTeX style support page is:
%  http://www.michaelshell.org/tex/ieeetran/bibtex/

\bibliographystyle{IEEEtran}
\bibliography{template}

\begin{IEEEbiography}[{\includegraphics[width=1in,height=1.5in,clip,keepaspectratio]{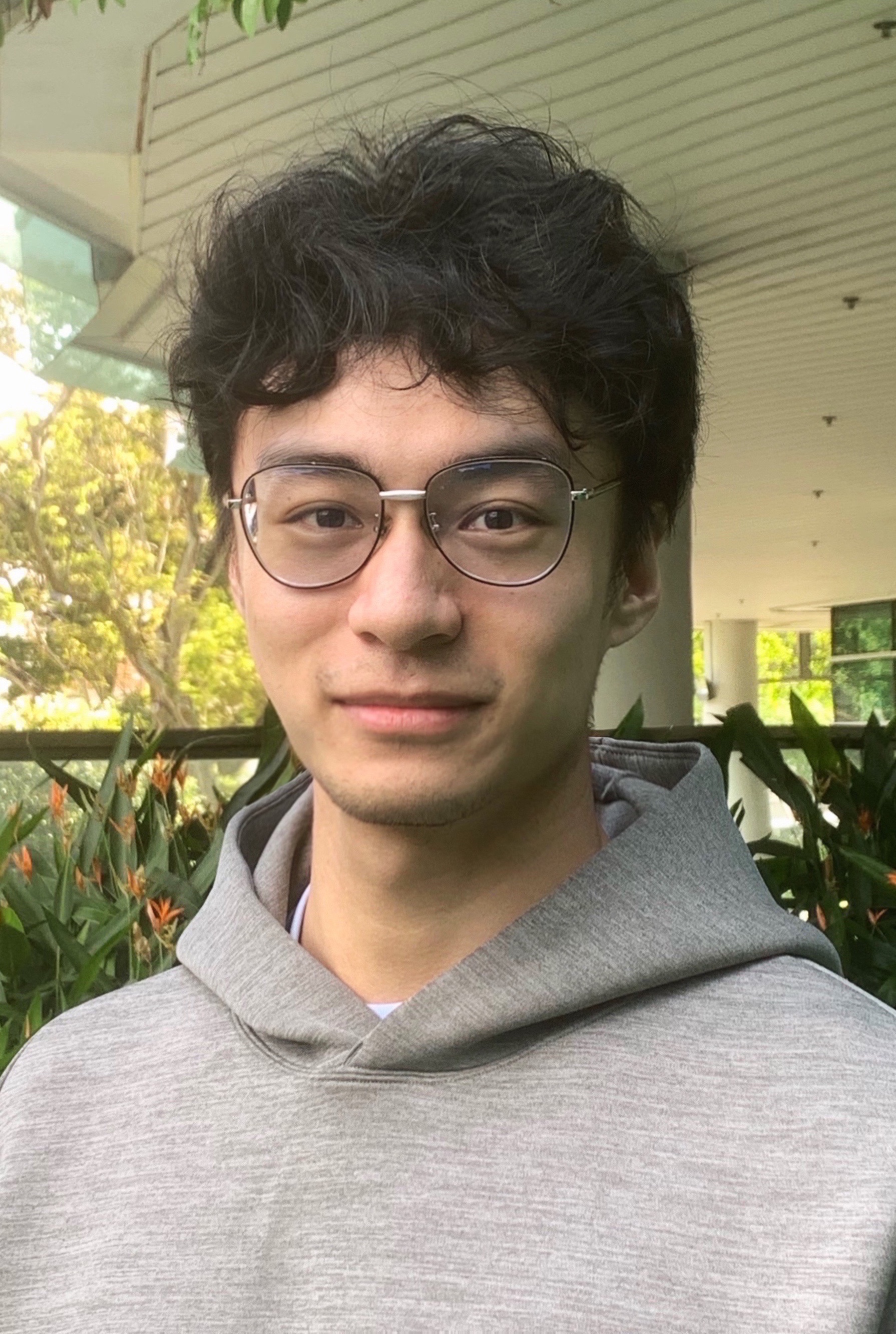}}]{Shaolun Ruan} is currently a Ph.D. candidate in School of Computing and Information Systems at Singapore Management University (SMU). 
His work focuses on developing novel graphical representations that enable a more effective and smoother analysis for humans using machines, leveraging the methods from Data Visualization and Human-computer Interaction.
He received his bachelor's degree from the University of Electronic Science and Technology of China (UESTC) in 2019.
% majoring in Information Security at the School of Computer Science and Engineering in 2019. 
For more information, kindly visit \url{https://shaolun-ruan.com/}.
\end{IEEEbiography}

\begin{IEEEbiography}[{\includegraphics[width=1in,height=1.5in,clip,keepaspectratio]{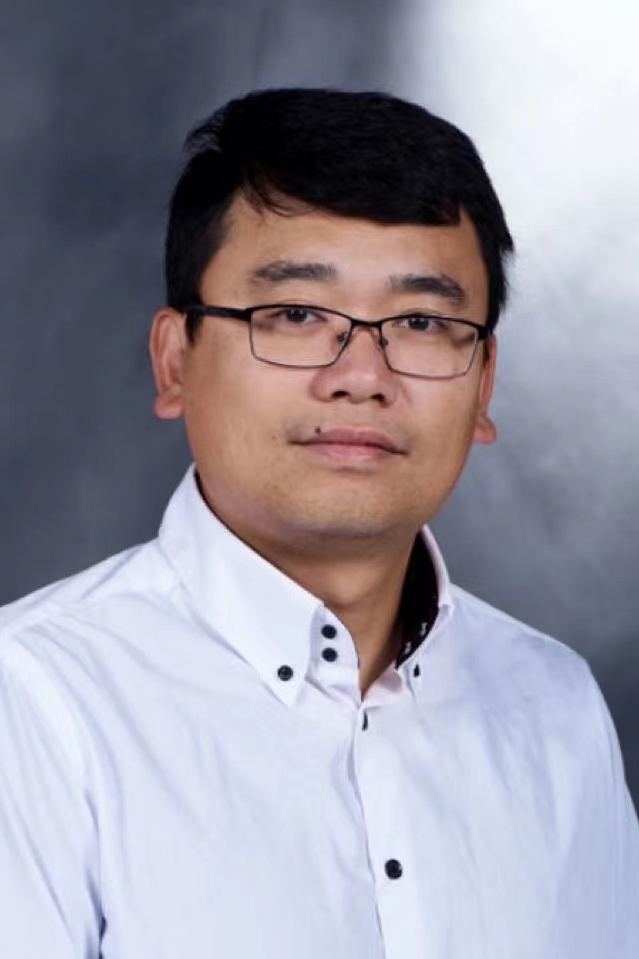}}]%
{Qiang Guan}
is an assistant professor in Department of Computer Science at Kent State University, Kent, Ohio. He is the direct of Green Ubiquitous Autonomous Networking System lab (GUANS). He is also a member of Brain Health Research Institute (BHRI) at Kent State University. He was a computer scientist in Data Science at Scale team at Los Alamos National Laboratory before joining KSU. His current research interests include HPC applications, quantum computing systems, HPC-Cloud hybrid system, virtual reality, and applications. 
% For more details, please refer to \url{http://www.guans.cs.kent.edu/}
\end{IEEEbiography}

\begin{IEEEbiography}[{\includegraphics[width=1in,height=1.5in,clip,keepaspectratio]{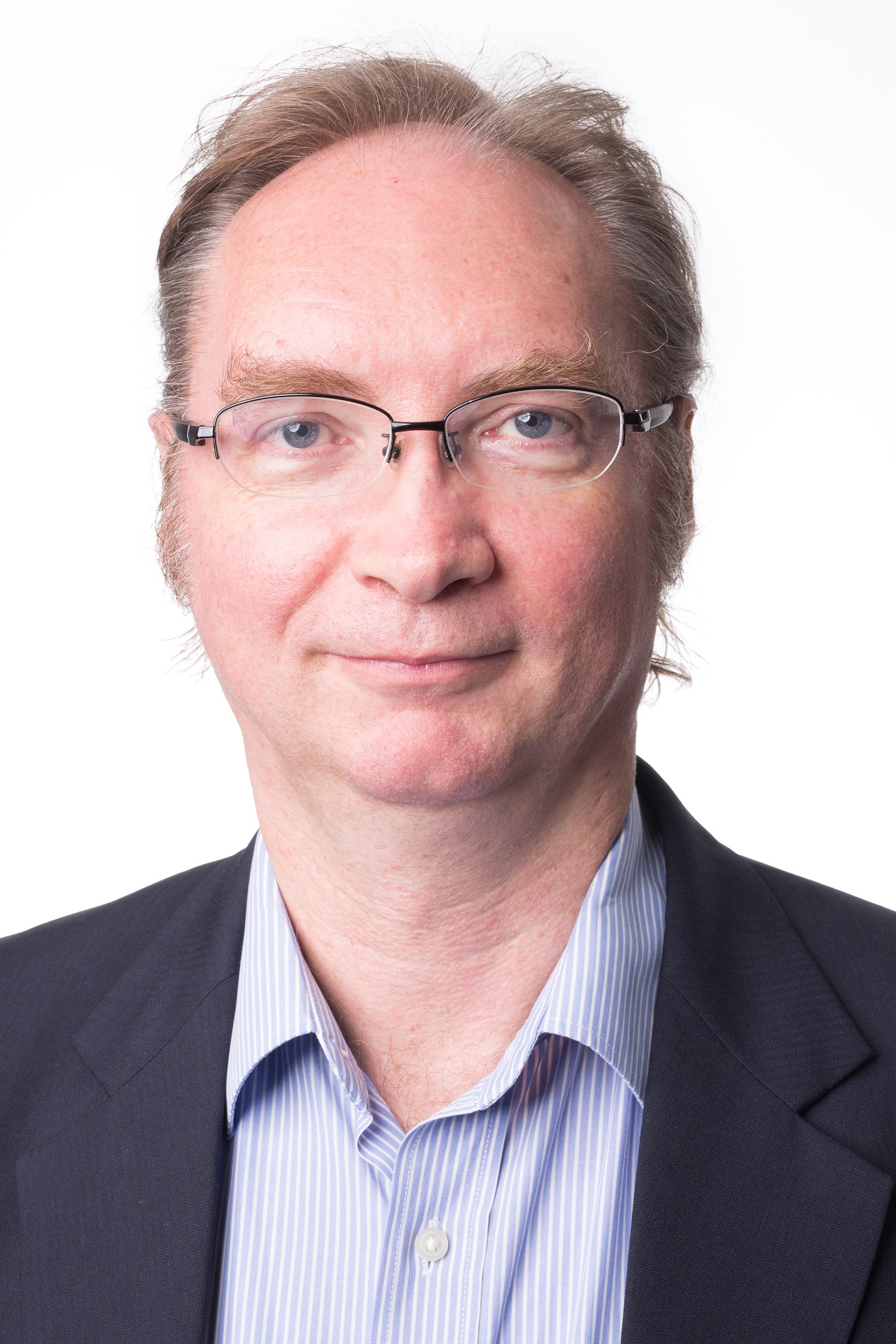}}]%
{Paul R. Griffin}
is in SMU working on disruptive technologies in FinTech as an Associate Professor. He gained a PhD at Imperial College in quantum physics and, prior to SMU, lead application development teams on global financial technology projects for over 15 years in the UK and Asia. Paul has been advising companies on disruptive technologies since 2014 and is now teaching and researching on blockchain and quantum computing as well as presenting at events, judging hackathons and moderating panel discussions.
\end{IEEEbiography}

\begin{IEEEbiography}[{\includegraphics[width=1in,height=1.5in,clip,keepaspectratio]{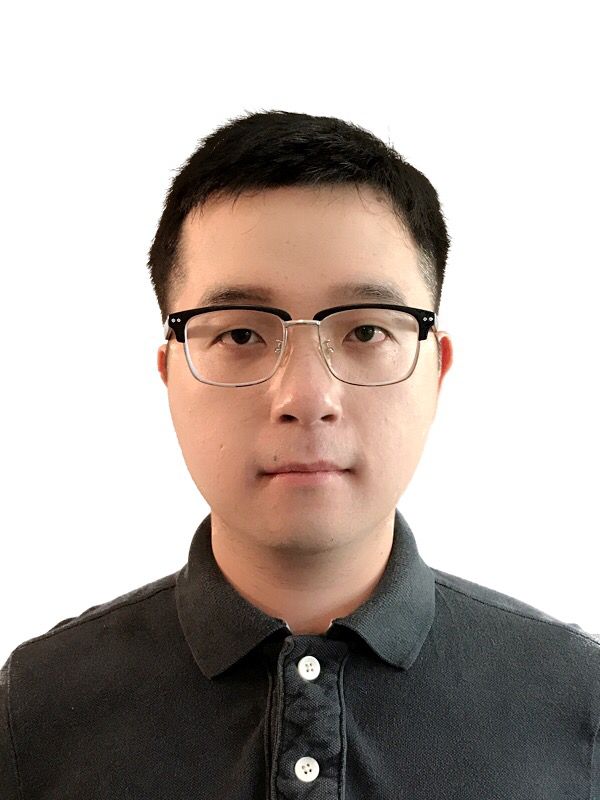}}]%
{Ying Mao} is an Associate Professor in the Department of Computer and
Information Science at Fordham University in the New York City. He
received his Ph.D. in Computer Science from the University of
Massachusetts Boston in 2016. He was a Fordham-IBM research fellow.
His research interests mainly focus on quantum systems, quantum deep
learning, quantum virtualization, quantum resource management,
data-intensive systems, and containerized applications.

\end{IEEEbiography}

\begin{IEEEbiography}[{\includegraphics[width=1in,height=1.5in,clip,keepaspectratio]{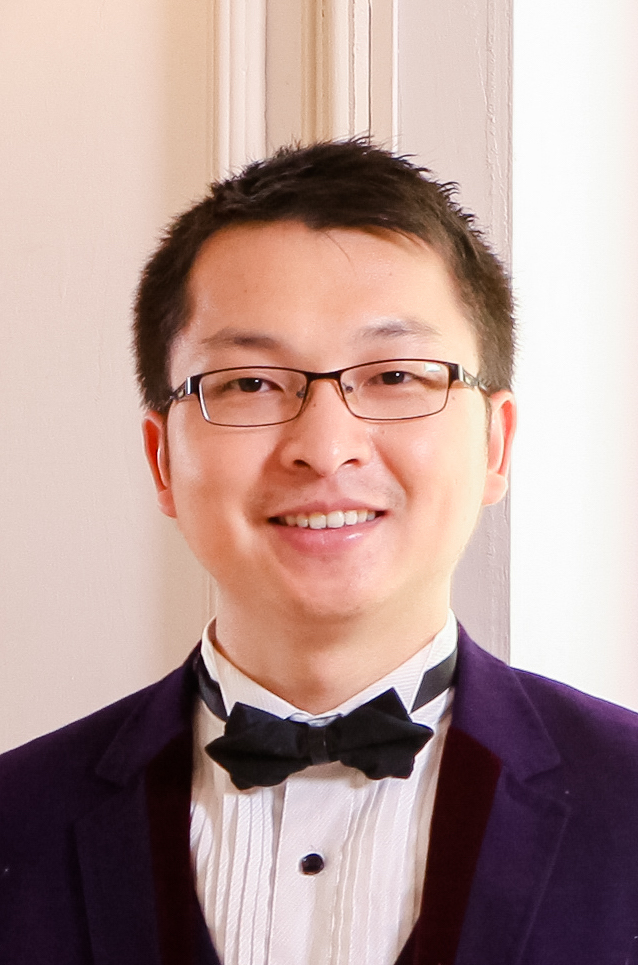}}]{Yong Wang} is currently an assistant professor in School of Computing and Information Systems at Singapore Management University. His research interests include data visualization, visual analytics and explainable machine learning.
He obtained his Ph.D. in Computer Science from Hong Kong University of Science and Technology in 2018. He received his B.E. and M.E. from Harbin Institute of Technology and Huazhong University of Science and Technology, respectively. For more details, please refer to \url{http://yong-wang.org}.
\end{IEEEbiography}

\vfill

% % Appendix

% \newpage
% \clearpage

% \appendices
% \input{sections/appendix}

\end{document}

%% file: sections/1-introduction.tex
\section{Introduction}

% \maketitle

Quantum computing has experienced remarkable advancements in recent years. 
\revise{The rapid growth in the quality and quantity of quantum computers by leading IT companies, such as IBM, Google and Amazon, are} making potential quantum advantages increasingly realistic for both 
% quantum computing has been undergoing impressive development ranging from 
theoretical quantum algorithms~\cite{toyama2013quantum, politi2009shor, van2006quantum, hallgren2007polynomial} and emerging applications~\cite{arute2019quantum,castelvecchi2017ibm,biamonte2017quantum,orus2019quantum,hassija2020present}. 
For example, quantum computing has shown its superior speedup on classical problems, such as \textit{Grover's algorithm} for unstructured search~\cite{toyama2013quantum}, and \textit{Shor's algorithm} for integer factoring~\cite{politi2009shor}.
% \yong{Pls double-check if it is a correct claim.}
%
%
Meanwhile, researchers began to explore the power of quantum computing in various applications, \revise{such as machine learning~\cite{biamonte2017quantum}, finance~\cite{orus2019quantum}, and chemistry~\cite{hassija2020present}}.
\revise{The quantum supremacy experiment by Google~\cite{arute2019quantum} has shown the potential advantage of quantum computers over their classical counterparts.}
% The recent milestone claimed by Google in quantum error correction~\cite{google2023suppressing} has further improved its chance of success.}
% \yong{Folks, pls check the claims here.}
%
%
% %
% \revise{\st{Also, Google has demonstrated a real advantage of quantum computers over their classical counterparts, known as the experiment of} \textit{quantum supremacy}~\cite{castelvecchi2017ibm}.} \yong{Is Ref. \cite{castelvecchi2017ibm} talking about Google? Be careful!} 
% More recently, a groundbreaking  milestone claimed by Google in quantum error correction~\cite{google2023suppressing}
% % making its immense commercial value even more realistic.
% has further improved its chance of success.

\begin{figure}[t]% specify a combination of t, b, p, or h for top, bottom, on its own page, or here
  \centering % avoid the use of \begin{center}...\end{center} and use \centering instead (more compact)
  \includegraphics[width=0.85\columnwidth]{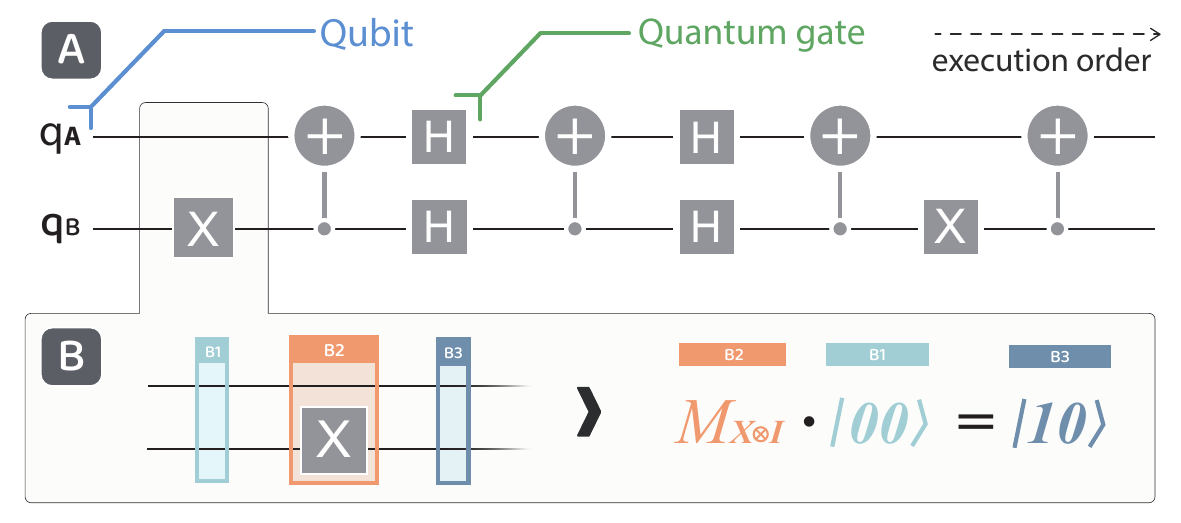}
  \caption{(A) An example quantum circuit reproduced from the prior work~\cite{ash2019qure}, \revise{which consists of qubits and quantum gates. (B) The intuitive illustration that shows how the matrix multiplication is performed given the quantum gate and initial quantum state.}}
  \label{fig:1}
\end{figure}

Building upon the proliferation of quantum computers, the number of people learning quantum computing has experienced rapid growth in recent years~\cite{piattini2020training}.
% and the demand of the quantum computing experts has become rather strong in IT companies~\cite{piattini2020training}.
However, prior research  has identified  that grasping abstract concepts in quantum computing remains challenging~\cite{lin2018quflow, williams2021qcvis}.
For example, \textit{quantum circuits}, the most fundamental routine to perform any quantum program,  lack the transparency and interpretability needed for easy comprehension~\cite{williams2021qcvis}. 
% Meanwhile, visualization has been proven to be an effective learning tool for educating people on obscure scientific concepts~\cite{vavra2011visualization, phillips2010visualization}.
Consequently, a graphical representation~\cite{feynman1985quantum} known as quantum circuit diagrams was proposed decades ago
% . This approach continues to be 
and \revise{has been widely used in research papers and textbooks on quantum computing.
% printable papers and textbooks throughout the quantum information literature.
}
% As depicted  in Fig. \ref{fig:1}\component{A}, 
% The quantum circuit diagram 
% provides 
% It can provide
% users with an overview of the architecture of a quantum circuit with wires and symbols.
% Despite its prevalence,
% it primarily illustrates the construction of a quantum circuit and 
% cannot reveal deep insights into the quantum circuit's behaviors.
Despite its prevalence,
it primarily overviews a quantum circuit and 
\revise{has limitations} in revealing deep insights into quantum circuits' behaviors.
% From a quantum circuit diagram like Fig. \ref{fig:1}\component{A}, it is difficult for quantum computing users, especially novice users, to understand the inherent functionality of each quantum gate and the final outcomes measured upon the quantum circuit's completion.
From a quantum circuit diagram, it is difficult for quantum computing \revise{developers and researchers} to understand the functionality of each quantum gate and the final measured probability of each basis state.
% \yong{Why change the sentence and remove the example diagram?}
% \shaolun{maybe someone comments it. I reopened the example.}
% Specifically, when examining a quantum circuit diagram like Fig. \ref{fig:1}\component{A}, the viewers may struggle to fully understand the inherent functionality of each quantum gate and the final outcomes measured upon the quantum circuit's completion.
% how the final measured probability is generated or in this quantum circuit.
% from the quantum circuit diagrams like Fig. \ref{fig:1}\component{A}.
For example, the viewers cannot inspect the quantum states' initial generation and further evolution or the functionality of each quantum gate from a quantum circuit diagram (\textit{e.g.}, Fig. \ref{fig:1}\component{A}). 
% Thus, this type of visualization struggles to provide users with intuitive insights into the inner workings of a quantum circuit.
% To this end, a new visualization approach supporting a better understanding of quantum circuits is urgently needed to facilitate the learning process. 
Thus, how to intuitively reveal the detailed inner workings of a quantum circuit still remains under-explored.
% and a new visualization approach for better interpretability of quantum circuits is urgently needed.
% to facilitate the learning process. 

% \yong{Need a further check}

However, it is non-trivial to fill this research gap.
According to our extensive literature survey~\cite{preskill2018quantum, lin2018quflow, hey1999quantum, bardin2021microwaves, wie2014bloch, steane1998quantum, sruan23} and close collaborations with six quantum computing experts,
the major challenges mainly come from
% two perspectives:
the counter-intuitive nature and intrinsic complexity of \textbf{\textit{quantum gate operations}} and \textbf{\textit{measured probability}} of quantum circuits.
First, the quantum gates are the fundamental and crucial operators to manipulate the state of qubits.
% Mao--- also the fundamental building blocks for any quantum algorithms. 
% , which makes it  for domain users to understand quantum gate operations in order to interpret
% quantum circuits~\cite{williams2021qcvis, lin2018quflow}.
% Qubits serve as the basic units for storing quantum states, while quantum gates act as fundamental operators that manipulate these states by affecting one or more qubits. 
% However, visualizing gate operations along a quantum circuit remains a challenge, as 
But quantum gate operations are essential matrix multiplications that are difficult to \revise{visualize} and explain.
% the transformation of quantum gates on qubits involves matrix multiplication, an abstract mathematical concept that is difficult to visualize and explain.
% Moreover, the matrix of quantum gates' transformation consists of complex numbers as dictated by quantum theory~\cite{narayanan1997introductory}. This counter-intuitive nature further complicates the visualization of quantum gate behavior.
What makes matters worse is the matrix transformations of quantum gates involves complex numbers
% as dictated by quantum theory
~\cite{narayanan1997introductory} that are counter-intuitive.
Second, the measured probability, determined by the quantum amplitudes of each basis state, is critical to understand the output of quantum circuits.
But users often do not possess a mathematical intuition regarding the underlying cause of each basis state's amplitude~\cite{sruan23}.
Also, quantum system states of multiple qubits can be entangled together rather than being a simple accumulation of multiple individual single-qubit states, and there will be $2^{N}$ possible \revise{basis} states if the qubit number is $N$, making it extremely challenging to visualize multi-qubit states and the corresponding measured probabilities in a limited space~\cite{sruan23}.

To address
% \revise{some of} 
the above challenges, 
% \yong{Note: the revision is declined. Pls update the cover letter.}
we propose \toolName{}, a visualization approach to enhance the interpretability of quantum circuits.
\toolName{} can intuitively explain the functionality of each quantum gate and the measured probability of each basis state for a given quantum circuit.
% To ensure the effectiveness of our visual designs, 
We follow a user-centered design process~\cite{munzner2009nested} by working closely with six domain experts in quantum computing for over five months.
% A well-designed preliminary interview is conducted to derive the design requirements, which guides the subsequent visual design for \toolName{}.
% Based on the derived feedback, we propose two workflows 
By summarizing the expert feedback,
we distilled design requirements in terms of two levels of analysis
- \emphasize{global} analysis and \emphasize{local} analysis.
% - to address the two aforementioned challenges.
For the global analysis, 
% the explainability of the qubits and quantum gates,
we propose three coordinated views to enhance the interpretability of quantum gates' operations:
a Probability Summary View summarizes the changes of all quantum states along a circuit (Fig. \ref{fig:case1}\component{A}),
a State Evolution View supports analyzing how quantum gates affect the evolution of multiple quantum states over time (Fig. \ref{fig:case1}\component{B}),
and a Qubit Explanation View further explains the quantum gates' effect from the view of the single qubit and its acting quantum gate (Fig. \ref{fig:case1}\component{C}).
% Particularly, the probability timeline chart
% The state evolution chart.
% Single-qubit explanation chart
For the local analysis, 
we propose a novel geometrical visualization \designName{} (Fig. \ref{fig:5}), which can visualize and explain the measured probability of numerous basis states based on amplitudes.
% support the multi-qubit state visualization, building upon the useful property of highlighting the probability of each quantum state. 
% Meanwhile, the design can also address the scalability issues by user interactions based on geometric characteristics.
% \designName{} is 
To evaluate the usefulness and effectiveness of \toolName{}, we present two case studies based on the famous quantum algorithms,
% quantum circuits with different quantum circuit architectures and qubit numbers, 
\textit{i.e.}, \textit{Grover's Algorithm} and \textit{Quantum Fourier Transform}.
We further conducted in-depth interviews with 12 domain experts with carefully designed tasks.
The results show that \toolName{} can effectively help developers and researchers better understand the behaviors of quantum circuits.

The major contributions of this paper can be summarized as follows:

\begin{itemize}
    \item We formulate the design requirements for improving the interpretability of quantum circuits by working closely with quantum computing experts.

    \item We introduce \toolName, an interactive visualization system to assist quantum computing users in intuitively understanding the behaviors of quantum circuits, including three coordinated views for global analysis and a novel design \designName\ for local analysis.

    % \item We propose a geometrical visual design, \designName, to support the intuitive explanation of how amplitudes affect the probability of basis states in the perspective of a specific quantum state.

    \item We present two case studies and in-depth user interviews with domain experts to demonstrate the effectiveness and usability of \toolName.
    
    % \item We make the \toolName\footnote{\systemURL} publicly available to ensure that quantum computing users can yield benefits from our study. We also publish the package of \designName\footnote{\designURL} to support a quick build of the quantum state visualization following the suggestions from domain experts.

\end{itemize}

To further benefit quantum computing developers and researchers, we have made our system \toolName{} publicly accessible online\footnote{\systemURL}. Also, we have published \designName{} as an independent NPM package\footnote{\designURL}.
% to facilitate the quick visualization of quantum states.

%% file: sections/2-related_work.tex
\section{Related Work}

Our work is relevant to prior research on 
visualization of quantum circuit evolution
and 
quantum state visualization.

\subsection{Quantum State Visualization}

Many existing approaches studied how to represent quantum states, the mathematical description of the state of a quantum system.
% modified by a quantum gate after each step of a quantum circuit.
We classify existing visual representations for quantum states based on whether the visualization is state vector-based or probability-aware.

\textbf{State vector-based approaches.}
The state vector-based approach aims to visualize the quantum amplitudes of quantum states.
\revise{The most widely-used representation in the quantum computing community is \textit{Bloch Sphere}~\cite{bloch1946nuclear}, which is integrated into many popular quantum computing SDKs like IBM Qiskit~\cite{ibmqiskit} and Google Cirq~\cite{googlecirq} to visualize quantum states.} Bloch Sphere leverages a point on the unit sphere to represent the quantum amplitude of a pure single-qubit state. 
\revise{Meanwhile, Bloch Sphere can also reflect two important visual effects, \textit{i.e.}, single-qubit rotation gates and statistical mixtures of pure states.}
Prior work has introduced various \revise{extensions of Bloch Sphere}~\cite{makela2010n, wie2020two, altepeter2009multiple}.
Also, many researchers have studied how to represent quantum states using 2D shapes.
Wille et al.~\cite{wille2021visualizing} visualized the components of state vectors using a tree-like design.
\revise{Several studies explored how to better visualize quantum states by enabling multi-qubit visualization, such as the stellar representation~\cite{bengtsson2017geometry} and the visualization based on multi-qubit Bloch vectors~\cite{algassert}.}
% Although Bloch Sphere is widely used by quantum computing users, several issues still exist.
However, several issues exist in the above visualization approaches.
% First, it cannot support multi-qubit state visualization~\cite{wie2014bloch, hooyberghsintroducing}, which is necessary as entangled quantum states, based on multiple qubits, are the basic ingredient of quantum computing.
First, \revise{
% for the encoding of these visualizations, the probabilities of all basis states cannot be directly compared by human eyes
these visualizations do not enable a direct comparison of the probabilities of basis states, making it hard for users to inspect
% the rationale of 
the measured probability.
Second, for 3D representations, they have been proven less effective than 2D counterparts when conducting precise measurements~\cite{tory2005visualization, FA20153D2D}.}

\textbf{Probability-aware approaches.}
Some prior work focused on improving the state vector-based approach by explicitly visualizing the measured probability based on the state vector representation.
For example, 
Galambos et al.~\cite{galambos2012visualizing} utilized a fractal representation of a multiple-qubit system via a set of rectangles.
Also, Chernega et al. studied several variants~\cite{chernega2019correlations, chernega2017triangle2} based on Triada of Malevich’s squares, which mapped the state vectors of a qubit onto the vertices of a triangle.
More recently, Ruan et al.~\cite{sruan23} introduced a 2D geometrical visualization to highlight the impact of the state vector on the probability.
% , preserving the nature of the probability measurement representation using multiple squares.

\revise{Similarly, Miller et al.~\cite{miller2021graphstatevis} proposed an interface with an embedded node-like graph to explain circuits and stabilizer groups, \revise{allowing the observation of updates of quantum states.}} 
% \yong{Isn't the description here conflicting the claim in the subsequent sentence?!}
Although the prior work can visualize the probability of quantum states, they still suffer from scalability issues. Most studies can only support the visualization of one qubit~\cite{chernega2019correlations, chernega2017triangle2} or two qubits~\cite{galambos2012visualizing, sruan23}, \revise{whereas most of the accessible quantum
computers are already exceeding this number of qubits. Therefore,} it is crucial to enable the inspection of more qubits.
Our work aims to support quantum state visualization with multiple qubits, while preserving the property of being probability aware.
% which is the focus of this paper.

\subsection{Visualization of Quantum Circuit Evolution}

We categorize existing work into two groups: depending on whether the proposed visualization technique is for a specific algorithm or general quantum circuits.

\textbf{Algorithm-specific visualization.}
Visualization approaches in this category often aims at a specific quantum algorithm without the generalizability for general quantum programs.
For example, Tao et al.~\cite{tao2017shorvis} utilized a Bloch Sphere and a disk-like design to portray the evolution of each quantum states along each step of \textit{Shor's algorithm}.
\revise{Karafyllidis et al.~\cite{karafyllidis2003visualization} studied how to visually explain the QFT algorithm by visualizing the changes of the probability of each quantum state, but
% has the limitations of the trace-back of 
it cannot support the trace-back analysis of basis states.} 
% \yong{Trace or trace back? What is the actual meaning your want to convey here?}
Meanwhile, \revise{two online platforms~\cite{qecvis, qecviss} enabled users to visually understand quantum states and quantum circuits in Quantum Error Correction, respectively.}
But it is challenging to extend to arbitrary quantum circuits, which significantly limits their benefits and impact.

\textbf{Generally-applicable visualization.}
% Generally-applicable methods mainly focus on the explanation of general quantum circuits.
Unlike algorithm-specific explainability, generally-applicable methods can be applied to arbitrary quantum circuits and thus are more flexible.
One common approach is leveraging measured probability to depict each step's behavior in a quantum circuit.
For example, Williams~\cite{williams2021qcvis} and Lin et al.~\cite{lin2018quflow}
showed
the probabilities of all possible states after each quantum gate to interpret the gate's functionality.
\revise{Wen et al.~\cite{wen2023quantivine} also studied to improve the scalability of large-scale circuits.}
% \revise{However, the vertical coordinates of their visualizations are based on qubits and basis states, making it challenging for users to inspect how quantum gates directly affect the probabilities of basis states.} 
% \yong{Why? It does not make sense to me.}
%
%
\revise{Lamy~\cite{lamy2019dynamic}
% \yong{For single author, you only need to mention his/her family name.}
studied how to reveal the gate effect by visualizing the change of quantum state in each step with a rainbow box design,
while preserving the display of phases.} 
% \yong{What is ``the feature of visualizing phases''?}
Moreover, \revise{Van de Wetering~\cite{van2020zx} proposed a graphical representation of a linear map between qubits.} 
% \yong{How is it extended? For what? Need a CONCRETE idea here.}
% visually interpret the feature of each quantum gate using the following quantum states' probabilities.
Another type of work focuses on explaining the noise in quantum circuits. For example, Ruan et al.~\cite{ruan2023isualization} introduced a visualization approach for the awareness of noise hidden in quantum computers and compiled quantum circuits.
% In addition to the research paper, \textit{Quirk}~\cite{quirk}, an online platform, also provides users with an interactive graphical interface to visually simulate the development of quantum circuits.
\revise{Meanwhile, Quirk~\cite{quirk} and Q-Sphere~\cite{qsphere} also enable users to interact with quantum circuits via a web-based platform.}
% Unlike the aforementioned methods, our work provides a generally-applicable visualization approach to explain 

While all the above methods focus on visualizing the quantum circuit evolution via the sequence of basis states' probability,
our work aims to depict the development of a quantum circuit by visualizing the basis states' relationship with a more effective visual channel, \textit{i.e.}, position~\cite{cleveland1987graphical}.
Also, \toolName{} uses Gate Explanation View and \designName{} to explicitly explain gate functionality with greater clarity.
% However, according to prior work~\cite{galambos2012visualizing, sruan23}, users look forward to finding the hidden rationales behind the probability change other than solely visualizing the high-level probabilities.
% Our work aims to visually explain the measured probability by the correlated amplitudes, enhancing users' confidence about the quantum circuit evolution.
% the state evolution analysis and quantum amplitude comparison, 

%% file: sections/3-background.tex
\section{Background}

This section introduces the background of quantum computing relevant to our study,
including 
% the basic blocks of quantum computing, 
quantum states and quantum circuits.

\subsection{Quantum State}

In quantum computing, quantum states are the mathematical entities that provide the probability of multiple basis states. 
% The true power of quantum computing derives from the exponentially increasing state space,
% as there will be $2^{N}$ basis states simultaneously for a specific quantum state with $N$ qubits~\cite{hey1999quantum, hughes2021quantum}.
Meanwhile, the actual calculation of gate operation can be represented as the \revise{matrix multiplication} of quantum states and gates (\textit{e.g.}, Fig. \ref{fig:1}\component{B}).
Recalling that for one qubit, the single-qubit state can be expressed as $\alpha\ket{0} + \beta\ket{1}$.
% \revise{The notation $\ket{\phi}$, 
% \yong{why ``in'' here?}
% namely Bra-ket notation~\cite{dirac_1939}, is a standard mathematical framework used frequently to represent quantum states.}
Generally, any quantum state with $n$ qubits can be represented as a linear combination of $2^{n}$ basis states:

\begin{equation}
\label{equation:2}
\alpha \cdot \ket{0 \cdots 00} + \beta \cdot \ket{0 	\cdots 01} + 	\cdots + \gamma \cdot \ket{1	\cdots 11},
\end{equation}

where the complex number $\alpha, \beta, \cdots \gamma$ are called quantum amplitudes (\textit{a.k.a.} amplitudes) which is used to describe the basis state (\textit{e.g.}, $\ket{0 \cdots 01}$) of a quantum state. An arbitrary amplitude (\textit{e.g.}, $\alpha$) can be expressed as a complex number: 

\begin{equation}
\label{equation:3}
\alpha = a + b \cdot i,
\end{equation}

where $a$ is the real part, and $b \cdot i$ is the imaginary part ($i$ is the imaginary unit).
Note that the amplitude of any quantum state can be used to determine the probability of measuring the corresponding basis state, which can be written as follows:

\begin{equation}
\label{equation:4}
Pr(\ket{0 \cdots 00}) = |\alpha|^{2} = |a|^{2} + |b|^{2}.
\end{equation}

Since the amplitudes of all basis states satisfy a normalization constraint that the sum of the probabilities of all basis states equals 1, thus all amplitudes satisfy $|\alpha|^2 + |\beta|^2 + \cdots + |\gamma|^2 = 1$.
\revise{Note that
we use the phrase ``measured probability'' in this paper to refer to the probability of a certain basis state if the quits were measured.}
% the phrase 
% ``measured probability'' refers to the probability of a certain basis state would be if qubits were measured.
% Also, we leverage the quantum simulator to access the measured probabilities before the final measurement at the end of the circuit execution.} 
% \yong{pls check if it is what you want to say.}

% \begin{equation}
% \label{equation:5}
% |\alpha|^2 + |\beta|^2 + \cdots + |\gamma|^2 = 1.
% \end{equation}

\subsection{Quantum Circuit}

\revise{Similar to classical circuits, quantum circuits describe how quantum algorithms can be decomposed into a sequence of physical gates (Fig. \ref{fig:1})}.
The manipulation of a quantum circuit can be represented as a calculation of unitary matrices~\cite{yao1993quantum}.
% For example, a quantum circuit of two qubits will be initialized from the start of the circuit, \textit{i.e.}, $\ket{00}$. 
% Then it will be manipulated by  a $4 \times 4$ unitary matrix~\cite{mottonen2004quantum}, \textit{i.e.}, NOT gate (\ref{fig:1}\Btwo{B\textsubscript{2}}).
% % which is.
% The final output of the NOT gate is the \revise{matrix multiplication} of the initial state $\ket{00}$ and transformation matrix $M_{X \otimes I}$, which yields $\ket{10}$.
In this paper, we refer to each manipulation module highlighted by the grey rectangle as a \textit{block}.
Thus, the execution of an arbitrary quantum circuit consists of the matrix calculation of a set of blocks, as illustrated in Fig. \ref{fig:case1}\component{A}.

Upon completing the final quantum gate, the execution result would be measured for the probability distribution of all basis states. Note that the intermediate quantum state after each gate's unitary transformation can be measured if the device is a quantum simulator~\cite{buluta2009quantum}. In contrast, only the final quantum state can be obtained for a real quantum computer due to the collapse of the quantum state upon measurement. 
% collapse of superposition?
\revise{Hence, for intermediate states, the visualization takes place in a ``god mode'' where the probabilities are known although the state is not actually measured.}

%% file: sections/4-informing_the_design.tex
\section{Design formation}

In this section, we first report the preliminary study with the design requirements distilled from the study.
We then introduce the dataset we used to fulfill the requirements.

\subsection{Preliminary Study}

% The primary goal of the preliminary study is to collect the design requirements faced in the routine tasks of quantum computing users.
Following the guideline~\cite{sedlmair2012design} of task abstractions for the design study, we designed the preliminary study as follows:

\textbf{Participants:}
The study involved six domain experts (\textbf{P1-6}) (6 males, $age_{mean}=36.5$, $age_{sd}=4.9$) from educational institutions and a national research laboratory.
Specifically, \textbf{P1-3} are professors from three different universities in Singapore and the U.S. \textbf{P4} is a research scientist from Pacific Northwest National Laboratory, and \textbf{P5-6} are two Ph.D. students whose research direction is quantum computing.
Among them, \textbf{P1-2} and \textbf{P5} are working on Quantum Machine Learning, while \textbf{P3-4} and \textbf{P6} study Quantum Systems, Quantum Chemistry and Quantum Error Modeling, respectively.
All the experts have an average of 6.8 years of research and development experience in quantum computing.

\textbf{Procedures:}
\revise{For} five months, we collaborated closely with the experts in quantum computing to conduct the preliminary study.
To ensure our approach was tailored to seamlessly fit into domain users' routine tasks, we \revise{divided} the whole procedure into two separate sessions.
First, we began the first session by performing one-on-one, semi-structured, hour-long interviews with all the domain experts.
During the interview, we posed carefully-crafted questions (see Appendix \ref{sec:appen4question}) relevant to the interpretability improvement of quantum circuits.
% , e.g., what is a quantum circuit's most crucial component to realizing a quantum algorithm's functionality?
For the second session, we summarized the initial design requirements and developed a low-fidelity prototype to meet the basic needs according to their feedback.
Next, we presented this prototype to the experts for iterative expert tests in the next three months.
They were then asked to explore the prototype freely and share their concerns and suggestions in a think-aloud manner; we then use their feedback to refine and improve the prototype accordingly.
% Throughout the study, we meticulously recorded observations and notes for each interview and discussion.

\subsection{Design Requirements}

We distilled the collected feedback from the preliminary study to inform our design.
Overall, we summarized users' general process as two levels of analysis, \textit{i.e.}, \emphasize{global} analysis and \emphasize{local} analysis.
Specifically, the global analysis (\textbf{R1-3}) aims to explain the effects of quantum gates from a high-level perspective, while the local analysis (\textbf{R4-6}) provides a more fine-grained explanation for the \textit{quantum states} by illustrating the rationale of the measured probability of each basis state.
% The combination of global and local analysis can provide both a high-level and fine-grained explanation of the quantum circuit, making it more thorough and rigorous for users to understand.
% Moreover, we also refined the requirements to enhance the \textbf{usability} for quantum computing users.

% For the \emphasize{global} analysis, users need to be aware of the functionality of quantum gates of quantum circuits from the following aspects:

\begin{itemize}

    \item[\textbf{R1}] \emphasize{global} \textbf{Provide an overall summary of the quantum circuit.}
    Five participants (\textbf{P1-4, P6}) emphasized the importance of providing users with a coarse-grained overview of the whole quantum circuit
    % . They all agreed with the idea of summarizing the evolution of quantum states 
    regarding the temporal changes of \textit{probabilities}, making it easier to interactively select the blocks of interest from a large number of gate operations.
    \textbf{P2} also mentioned the necessity to break the \textit{blocks} into a linear sequence of the individual gate operation, namely \textit{steps}, to illustrate the effect of each quantum gate better.

    \item[\textbf{R2}] \emphasize{global} \textbf{Explain the effect of quantum gates visually.}
    All participants (\textbf{P1-6}) strongly suggested that the visual designs should focus on the detailed explanation of the most basic ingredients (\textit{i.e.}, quantum gates).
    Specifically, three participants (\textbf{P1, P5-6}) encouraged us to utilize the basis states to depict the evolution of quantum states. Meanwhile, three experts (\textbf{P2-4}) also expressed the need to \textit{``visualize the quantum gate's effect via comparing how the \textit{amplitudes} change the measured probability before and after the quantum gate.''}
    % how they change the \textit{amplitudes} of quantum states.

    \item[\textbf{R3}] \emphasize{global} \textbf{Support the trace-back analysis of quantum states.}
    Three participants (\textbf{P1, P3, P5}) expected the approach to enable the trace-back analysis of quantum states. They all confirmed that it is significant to visually reveal how a specific quantum state was generated from the beginning of the quantum circuit. 
    % \textit{``I have no idea about how a quantum state is formed and by what kind of gate operations before\revise{,}''} \textbf{P1} commented, \textit{``I hope it can inform me of its evolution intuitively.''}
    Moreover, \textbf{P3} emphasized that the intuitive visualization of the original quantum circuit can significantly flatten the learning curves for domain users.
    
\end{itemize}

% For \emphasize{local} analysis, the following requirements are crucial for visualizing quantum states: 

\begin{itemize}

    \item[\textbf{R4}] \emphasize{local} \textbf{Explain the probability of basis states visually.}
    All participants (\textbf{P1-6}) confirmed that it would \revise{significantly} help to inform users of each basis state's probability change, enhancing their confidence \revise{in} understanding the effects of the quantum gates.
    In particular, four participants (\textbf{P1-3, P5}) emphasized the importance of visually correlating the amplitudes and probabilities other than by a set of individual visualizations (\textit{e.g.}, several bar charts), because they believed that the explicit and correlated visual channels could intuitively highlight how amplitudes determined the measured probabilities.

    \item[\textbf{R5}] \emphasize{local} \textbf{Support the visualization of multi-qubit quantum states.}
    According to the suggestions from four participants (\textbf{P1, P3, P5-6}), the most common visualization for quantum states, \textit{i.e.}, Bloch Sphere, cannot support multi-qubit state visualization.
    \textbf{P3} commented that this issue is unacceptable because the real power of quantum computing, \textit{i.e.}, entanglement, requires multiple qubits.
    \textbf{P6} also said \textit{``I really hope there exists a visual representation to make the multi-qubit state more intuitive.''}
    % regarding even hundreds of basis states,

    \item[\textbf{R6}] \emphasize{local} \textbf{Address the issues of visual scalability. }
    \textbf{P1} and \textbf{P3} pointed out the issue of visual scalability.
    % Specifically, \textbf{P1} emphasized that the scalability issues is a typical quantum-specific problem that needs to be addressed.
    \revise{Specifically, P1 emphasized that scalability issues are typical quantum-specific problems that need to be addressed.}
    Also,  \textbf{P1} commented
    % especially for those applications that need many qubits. 
   \textit{ ``There will be a substantial quantity of basis states in the common cases.''}
    \textbf{P3} also comments that visualizing \revise{many} basis states is a complex task, given the requirement to display both the probability and amplitudes of each basis state concurrently.

\end{itemize}

% For the overall visual designs, domain experts also highlight the importance of good usability for a visual analytics system when explaining quantum circuits.

% \begin{itemize}
%   % \setlength\itemsep{0pt}    
    
%     \item[\textbf{R7}] \textbf{Enable a flexible user interactions.}
%     Three participants (\textbf{P2, P4-5}) mentioned that the system should enable rich interactions to facilitate the on-demand analysis of users.
%     % , considering they rarely have a data visualization background.
%     Furthermore, \textbf{P1} also suggested that the system should be published to benefit more users in quantum computing community.

% \end{itemize}

% \shaolun{Delete to here}

\subsection{Dataset}

Building upon the above design requirements, we developed the system \toolName{} based on \textit{Qiskit}~\cite{qiskit}, which is an open-source framework for the implementation of quantum circuits.
We utilized a quantum simulator, \textit{i.e.}, \textit{AerSimulator}~\cite{aer_sim}, to extract quantum states.
% \toolName{} takes the structured data of the quantum circuit as input. 
The raw dataset extracted contains the properties of the quantum circuit: the sequence and implementation of quantum gates on the individual qubits, the state vectors of the quantum states over each step, and the transformation matrices of the quantum gates.

Next, to obtain the probability of each basis state, we \revise{leveraged} Equation \ref{equation:4} to calculate the amplitudes from the quantum state's state vector. 
Also, we \revise{decomposed} the matrix of state vector by rows to extract all basis states of a quantum state, making it available for analysis of the trajectory of the quantum states (see Appendix \ref{sec:appen4matrixDecom}).
Furthermore, based on the principle of unitary transformation~\cite{torch_quantum}, we \revise{deconstructed} each block (Fig. \ref{fig:case2}\component{C}) into multiple \textit{steps} (Fig. \ref{fig:case2}\component{B}) to better clarify the workflow of a quantum circuit.

%% file: sections/5-method.tex
\section{\toolName}

We proposed \toolName, an interactive visualization system to enhance the interpretability of quantum circuits.
% To achieve its availability for domain users (\textbf{R7}), the system was deployed online and can be accessed via the URL: \systemURL.  
% Fig. \ref{fig:2} illustrates the architecture of \toolName, which 
The architecture of \toolName{}
consists of three tightly-connected modules:
(1) data storage module, (2) data processing module, and (3) visualization module, as shown in Fig. \ref{fig:2}.
In particular, the data storage module stores all raw input data of the original quantum circuit.
% in respect of the sequence and implementation of quantum gates on the individual qubits, and the records of quantum states over each block.
The data processing module supports the data preparation procedure before visualization, including the probability calculation of each quantum state, the decomposition of state vectors for state evolution analysis, and the generation of the transformation representation based on the qubit states.
The visualization module reveals insights hidden in the quantum circuits,
where three views (\textit{i.e.}, Probability Summary View, State Evolution View, and \revise{Gate Explanation View}) are applied for the \emphasize{global} analysis and the novel design (\textit{i.e.}, \designName) is used for the \emphasize{local} analysis.
Furthermore, we also implement an original quantum circuit (Fig. \ref{fig:case1}\component{D}), enabling domain users to efficiently conduct the comparative analysis with our visual designs. The system interface of \toolName\ is shown in Appendix \ref{sec:appen4system}.

% \yong{Reach here.}

\definecolor{module_storage}{RGB}{159, 168, 69}
\definecolor{module_processing}{RGB}{159, 183, 172}
\definecolor{module_visualization}{RGB}{178, 78, 50}

\begin{figure}[t]% specify a combination of t, b, p, or h for top, bottom, on its own page, or here
  \centering % avoid the use of \begin{center}...\end{center} and use \centering instead (more compact)
  \includegraphics[width=0.95\columnwidth]{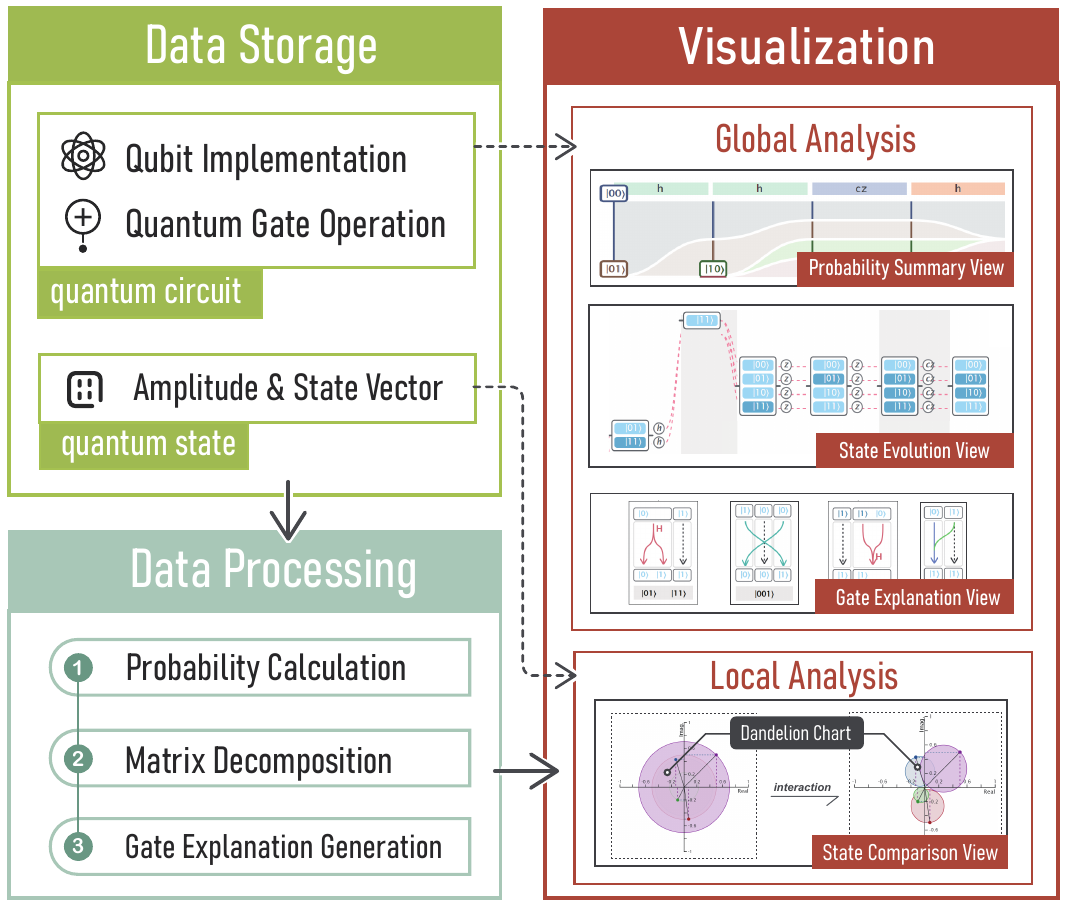}
  \caption{The system architecture of \toolName\ consists of a \textcolor{module_storage}{\textbf{data storage}} module, a \textcolor{module_processing}{\textbf{data processing}} module, and a \textcolor{module_visualization}{\textbf{visualization}} module.}
  \label{fig:2}
\end{figure}

\subsection{Probability Summary View}

\begin{figure}[t]% specify a combination of t, b, p, or h for top, bottom, on its own page, or here
  \centering % avoid the use of \begin{center}...\end{center} and use \centering instead (more compact)
  \includegraphics[width=0.9\columnwidth
  ]{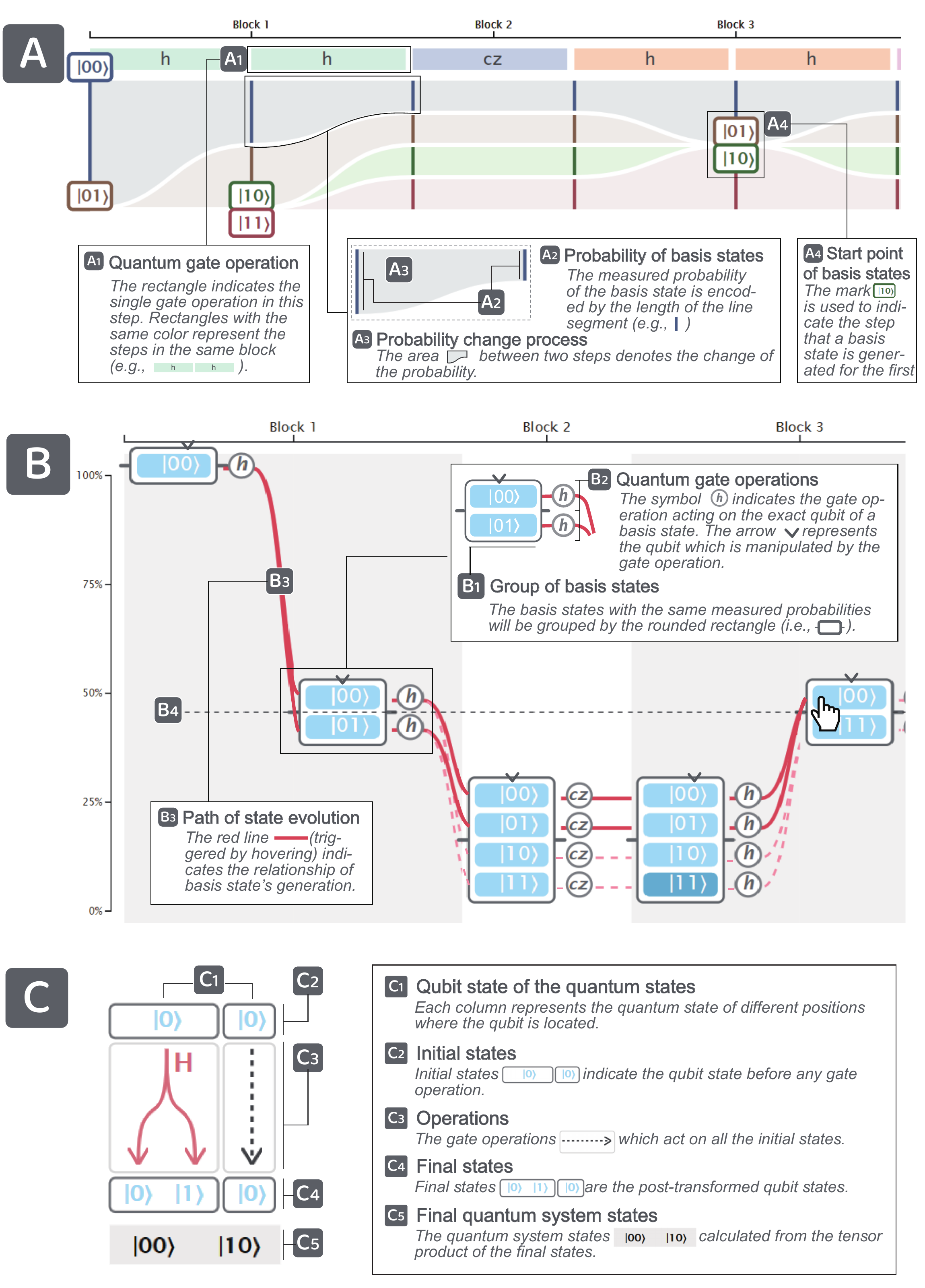}
  \caption{
 The three coordinated views in \toolName\ for \protect \emphasize{global} analysis.
 (A) Probability Summary View summarizes a quantum circuit via all basis states' temporal change of probabilities.
 (B) State Evolution View supports a fine-grained analysis of the basis states' evolution across each step.
 (C) \revise{Gate Explanation View} visually explains the effect of quantum \revise{gates} from the perspective of the qubit state.
  }
  \label{fig:4}
\end{figure}

We propose the Probability Summary View (Fig. \ref{fig:4}\component{A}) to provide an intuitive summary of the quantum circuit in terms of probability changes of basis states over each step (\textbf{R1}).
% We visualize the measured probability of each basis state that quantum computing users are most familiar with in practice.
% , to support a coarse-grained overview of the behavior of quantum circuits.
% Informed by \textbf{R1}, we first decompose the block into a sequence of steps to indicate the transformation of a quantum gate more explicitly.
% Thus, each step only indicates a single gate operation.
% , as illustrated in Fig. xxx.
% Building upon the decomposition of blocks, w
We leverage the stacked area chart to portray the basis state's measured probability on each step, where the length of line segments encodes the probability (Fig. \ref{fig:4}\subcomponent{A\textsubscript{2}}).
Specifically, we use a set of line segments arranged vertically to reveal the probability proportion at each step.
The total vertical length of all line segments at each step is a constant as the sum of all basis states' probabilities will always be 1.0.
% Recall that for a certain quantum state, the sum of all basis states' probabilities will always be 1 (\textit{i.e.}, Equation \ref{equation:5}); thus, .
Also, we utilize the area (Fig. \ref{fig:4}\subcomponent{A\textsubscript{3}}) to highlight the probability change of each basis state between steps.
% along the entire Probability Summary View. 
Moreover,
% \revise{to improve the comparison with the original quantum circuit}, \yong{Why? What do you want to convey here?}
%
%
we use a set of rectangles to denote the hierarchy of blocks and steps, where the rectangles in the same color are in a common block (Fig. \ref{fig:4}\subcomponent{A\textsubscript{1}}).
\revise{Note that the order of qubit labeling in the annotation is from left to right, while the qubit order in the view of original quantum circuit is from bottom to top.}
% \yong{Why do you add this sentence? What do you want to say here? If it is only for clarifying the labels, you can explain it in the figure caption instead of the paper.}
% Furthermore, according to the requirements collected from the preliminary interview, it is beneficial
Furthermore, we append the annotations (\textit{e.g.}, ~\raisebox{-2pt}{\annotation}) at the left-most area (Fig. \ref{fig:4}\subcomponent{A\textsubscript{4}}) to depict the creation of a basis state.
% a basis state generates for the first time.
% \textbf{User Interaction. }
To enable the drill-down analysis from the summary of the quantum circuit (\textbf{R1}), users can interactively brush the steps of interest in Probability Summary View.
% e.g., the functionality block \textit{Oracle} in \textit{Grover's Algorithm}~\cite{grover1996fast}.
% via brushing the preferred area.
% Probability Summary View will use the entire steps as default before the brushing by users.

% \textit{Justification. }
% We tried to use a multiple-line chart to represent the probability change of each basis state along the circuit，
% % (see \ref{subsec:view1}),
% where each line represents the basis state of the quantum system state.
% However, this method did not scale well due to the issue of edge crossing and overlapping.
% Additionally, the design did not sufficiently reveal the proportion of each basis state's probability.
% Thus, we finally leveraged the stacked area chart to enable users to trace the entities of basis states more intuitively.

% \shaolun{reach here}

\subsection{State Evolution View}
% \revise{
% While preserving the nature of the overall summary of quantum circuits (\textbf{R1}),} 
The State Evolution View (Fig. \ref{fig:4}\component{B}) enables a drill-down analysis
% from selecting steps of interest selected in Probability Summary View; it provides a fine-grained analysis 
of the evolution of quantum states
such as the separation and merging of basis states for Hadamard gates~\cite{aharonov2003simple} (\textbf{R2}).
% Specifically, State Evolution View illustrates the transformation of quantum gates.
The design also supports the trace-back analysis (\textbf{R3}), making users aware of how a basis state was generated and further transformed by quantum gates.

We visualize the evolution of all the basis states using a graph-like design.
\revise{Due to the consistency of the encoding of the horizontal axis, State Evolution View can also enable users to better compare with the Probability Summary View(\textbf{R1}).}
The horizontal coordinate indicates the steps of the quantum circuit, while the vertical coordinate represents the basis state's measured probability.
We use rounded rectangles to represent the entity of the basis state.
Meanwhile, those basis states with the same probability are grouped by the outer rectangle (\textit{i.e.},  ~\raisebox{-2pt}{\hub}), as shown in Fig. \ref{fig:4}\subcomponent{B\textsubscript{1}},
% We then apply the  to group (),
where the outer rectangles' short line segments refer to each group's measured probability.
Moreover, we encode the evolving relationship between the two steps using pink dotted lines.
To indicate the gate operation, we use a symbol with the acronym inside after each basis state (Fig. \ref{fig:4}\subcomponent{B\textsubscript{2}}); we then mark the qubit that the quantum gate acts on by the arrows.
Note that the rectangles will be colored in light blue~\lightblueblock\ if the basis state's real part is positive; otherwise, it will be colored in blue~\blueblock.
% To further empower the comparison with the original circuit, we differentiate the steps belonging to different blocks using two background colors grey and white.
% alternately color the background of the steps belonging to the same block the same using grey and white.
% \textbf{User interaction. }
We enable flexible interactions to enhance the usability of the system for users within the domain \revise{(\textbf{R3})}.
Precisely, users can hover over the specific state to analyze the evolution path highlighted in red lines (Fig. \ref{fig:4}\subcomponent{B\textsubscript{3}}).
% Also, for a more accurate inspection, a dotted reference line will display when hovering over the group of states (Fig. Fig. \ref{fig:4}\subcomponent{B\textsubscript{4}}).
% Furthermore, users can trigger the rendering of the embedded \designName{} by clicking on the quantum gates of interest.

\revise{\textbf{\textit{Justification.}} Prior work has also studied to explain quantum circuits by visualizing measured probability.
Lin et al.~\cite{lin2018quflow} and Karafyllidis et al.~\cite{karafyllidis2003visualization} studied how to explain the behaviors of the overall quantum circuit using the encoding of color.
Williams~\cite{williams2021qcvis} utilized the length to indicate the measured probability of single qubits.
In \toolName, we use the vertical position of the basis state instead of the encoding of color or length since the position has been proven a more effective visual channel for human perception~\cite{cleveland1987graphical}.
}

\begin{figure*}[t]% specify a combination of t, b, p, or h for top, bottom, on its own page, or here
  \centering % avoid the use of \begin{center}...\end{center} and use \centering instead (more compact)
  \includegraphics[width=0.9\linewidth
  ]{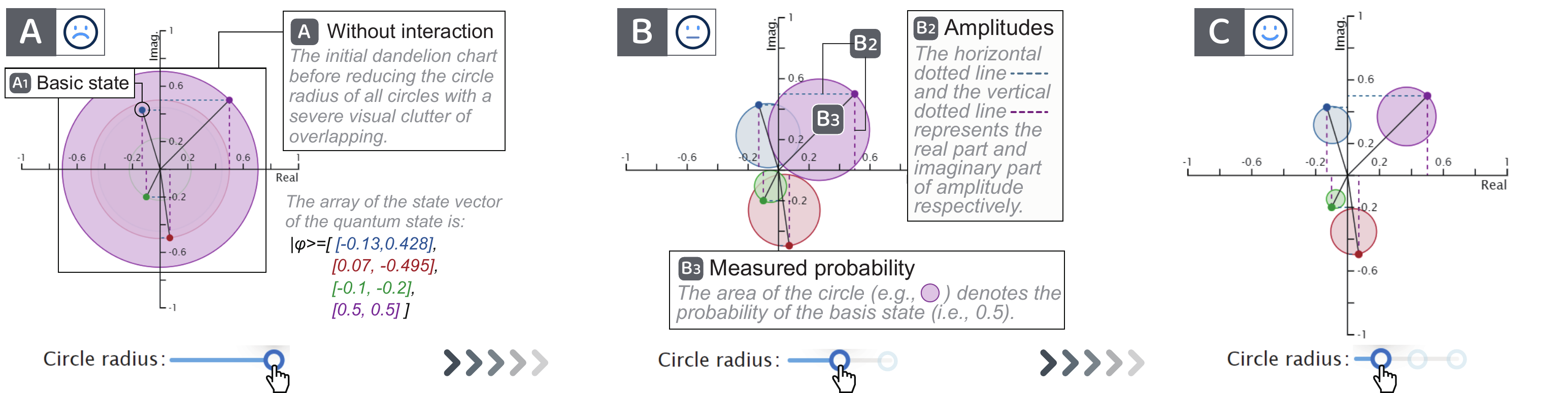}
  \caption{
The \designName\ embedded in Probability Explanation View.
(A) The \designName\ before interaction with all circles overlapped with each other.
(B) The \designName\ after reducing the area of all circles by a factor of \textbf{0.5}, where the visual clutter is mitigated slightly.
(C) The \designName\ after reducing the area of all circles by a factor of \textbf{0.25}, where all circles are completely separated apart and can be compared clearly.
  }
  \label{fig:5}
\end{figure*}

% \textit{Justification. }
% We considered the other two design alternatives before finalizing the current visual design (see \ref{subsec:view2}).
% First, we use the layout of a tree diagram to arrange all basis states. However, the expert \textbf{P1} pointed out that he cannot find any patterns regarding the gate's effect on the states since all states are located evenly at each step.
% We then came up with another design that visualizes the probability of each basis state using the opacity of the corresponding rectangles and depicts the state evolution by the dotted lines. 
% Despite intuition, the design does not scale well due to the severe line overlapping between two adjacent steps.
% Last, we propose the current design to address all these challenges.

\subsection{\revise{Gate Explanation View}}

The \revise{Gate Explanation View} (Fig. \ref{fig:4}\component{C}) aims to allow users to understand a gate operation based on the qubit state (\textbf{R2}).
We first deconstruct the quantum system states (\textit{e.g.}, $\ket{01}$) into qubit states (\textit{e.g.}, $\ket{0}$ and $\ket{1}$);
we then visualize the explanation via a table-like design. 
% \revise{We group it into the \textit{global} category since it is actually proposed to explain the quantum gates, which belong to the components of the quantum circuits.}

We define an arbitrary transformation as three parts, \textit{i.e.}, the initial state, operation, and the final state;
we then represent the three parts with the table's first, second, and third row, respectively.
The column denotes each qubit in the original basis state. 
Meanwhile, we apply various colored lines (\textit{e.g.}, ~\raisebox{-2pt}{\arrow} for Hadamard gates) to represent the operation of quantum gates acting on the individual qubits (Fig. \ref{fig:4}\subcomponent{C\textsubscript{3}}).
Note that the operation will be represented as the dotted grey line if no quantum gate acts on a qubit.
% The list of all the implemented representations of gate operations can be found in Appendix \ref{sec:appen4gate}.
For example, as shown in Fig. \ref{fig:4}\component{C}, assume there is a basis state ($\ket{{\color{blue} 0}0}$), the post-processed initial state is {\color{blue} $\ket{0}$} and $\ket{0}$.
After the Hadamard gate on the first qubit {\color{blue} $\ket{0}$}, the first qubit converts to a state in superposition, \textit{i.e.}, {\color{blue} $\ket{0}$} and {\color{blue} $\ket{1}$} each with a probability of 0.5, while the second qubit keeps as it is, \textit{i.e.}, $\ket{0}$.
Thus, the final state will be $\ket{{\color{blue} 0}0}$ and $\ket{{\color{blue} 1}0}$.
% (${\color{blue} \ket{1}} \otimes \ket{0}$)

% \textit{Justification. }
% We also came up with other design alternatives (see \ref{subsec:view3}),
% % to portray the transformation regarding qubits.
% % Fig.  shows the obsolete design 
% where a set of rectangles are used to represent the qubit states.
% % each rectangle represents the qubit state.
% The rectangle will be colored in white if the quantum state is $\ket{0}$. Otherwise, it will be colored in black for state $\ket{1}$.
% Also, the superposition is represented by the rectangle in half black and half white (Annotation \subcomponent{D\textsubscript{1}}).
% For the direction of rectangles, those placed vertically denote the initial states, while those placed horizontally depict the final states.
% However, this visual design is not preferred as it only depicts the initial and final state without explicitly visualizing the gate operations.
% % cannot be  visualized 
% Thus, we propose the current design, which intuitively highlights the gate operations process, making it easier for learners to understand the abstract transformations.

\subsection{Dandelion Chart}

To enable the explanation of measured probability (\textit{i.e.}, \emphasize{local} analysis), we propose \designName, a novel geometrical representation to visually explain the measured probabilities of basis states (Fig. \ref{fig:5}).
According to the quantum theory, we encode the amplitudes by 2D shapes to visualize arbitrary quantum states, including multi-qubit states (\textbf{R5}).
We also visually correlate the probability with the corresponding amplitudes based on geometry principles to explicitly explain the measured probability of basis states (\textbf{R4}).
Moreover, \designName\ allows users to mitigate the visual clutter of numerous basis states via a geometry-based approach (\textbf{R6}).
The \designName\ is incorporated into \toolName{} to facilitate the comparison between two quantum states before and after a gate operation, as shown in Fig. \ref{fig:case2}\subcomponent{B\textsubscript{4-7}}.
% To guarantee the availability of \designName, we implemented the approach with web-based SVG elements. 
% Domain users can build \designName\ instantly by downloading the package via the URL \designURL\ and simply call the function. 

% \subsubsection{\DesignName}

\textbf{Amplitudes encoding. }
To visually represent a quantum state and the respective basis states, 
we leverage amplitudes of quantum states as they are the basic components of a specific quantum state~\cite{deutsch1985quantum, valiev2005quantum}.
Recall that the amplitude of each state is intrinsically a complex number, consisting of a real and imaginary part, as illustrated by Equation \ref{equation:3}.
For each quantum state, we first apply a Cartesian coordinate system to represent the series of its amplitudes of each basis state based on Equation \ref{equation:2}, where the x-axis encodes the real part, and the y-axis encodes the imaginary part.
Thus, all the basis states of a quantum state are visualized as a set of points, as shown in Fig. \ref{fig:5}\subcomponent{A\textsubscript{1}}.
To further highlight amplitudes, the absolute values of real and imaginary parts are encoded by perpendicular lines in green and red from a point to the y-axis and x-axis (Fig. \ref{fig:5}\subcomponent{B\textsubscript{2}}).
Furthermore, we visualize the line connecting the point to the system's origin to highlight its position.

\textbf{Probability explanation.}
According to Equation \ref{equation:4}, the measured probability of each basis state can be calculated by the real and imaginary parts of the amplitudes.
Meanwhile, based on geometry principles, the circle's area can be calculated using the radius, which is equal to the distance between the basis states' points and the origin of the system:

\begin{equation}
\label{equation:6}
S_{circle} = \pi \cdot (|a|^{2} + |b|^{2}),
\end{equation}

where $a$ and $b$ are the real and imaginary parts of the amplitude.
Thus, building on the Equations \ref{equation:4} and \ref{equation:6}, we conclude that the area of the circle can represent the measured probability of a basis state as the area of the circle is proportional to the measured probability, as shown in Fig. \ref{fig:5}\subcomponent{B\textsubscript{3}}.
By this means, users are allowed to visualize the probability of the basis state in terms of their corresponding amplitudes indicated by the x- and y-coordinates of the points. 
However, there can exist a severe overlap between the circles (Fig. \ref{fig:5}\component{A}).
% We further address the visual clutter between different basis states.

% \subsubsection{Visual Clutter Mitigation}
\textbf{\revise{Visual clutter mitigation. }}
We mitigate the visual clutter by scaling the area of circles through user interaction.
By this means, all circles can be separated apart by decreasing all circles' radii, like the process from Fig. \ref{fig:5}\component{A} to Fig. \ref{fig:5}\component{C}, while preserving the nature of reflection of the state's probability using amplitudes.

% Recall that a basis state can be visualized as a pair consisting of a point and a circle, with the length of the line connecting from the point to the system's origin as the radius.
Specifically, if the radii of all the circles are reduced with the same factor $k$ while keeping the point on the edge of the circle.
% like the process from Fig. \ref{fig:5}\component{A} to Fig. \ref{fig:5}\component{C}.
Then the area of the circles satisfies the following equation:

\begin{equation}
\label{equation:7}
S^\prime_{circle} = \pi \cdot k^{2} \cdot (|a|^{2} + |b|^{2}),
\end{equation}

where $k \in [0,1]$ is the factor for shrinking the area of circles.
% Hence, the circles of various circles can be separated by decreasing all circles' radii, as shown in Fig. \ref{fig:5}\component{C}.
Meanwhile, based on Equations \ref{equation:4} and \ref{equation:7}, then the area of the circle is still proportional to the measured probability due to the constant factor $k$.
This finding means that users can scale the area of circles freely to mitigate the overlap while preserving the property of the representation of probabilities by the circles.
Hence, \designName\ can support probability explanations regarding amplitudes of the basis state through the user interaction of scaling the circles' radii.
We name the design as ``\designName'' due to the dandelion metaphor for each basis state like each entity in Fig. \ref{fig:5}\component{C}.

\revise{\textbf{\textit{Justification.}} The prior work by Lamy~\cite{lamy2019dynamic} also utilized the rectangle area to facilitate the measured probability analysis as well as the portray of entanglement and phase. 
Our novel design \designName, however, can \textit{explain} the measured probability regarding the amplitudes by the location and the corresponding circle area, while preserving the capability of phase and entanglement representation.
Specifically, building upon Cartesian coordinates, \designName{} encodes the probability by circle areas and further explains it by the location of the points based on the quantum mechanism constraints.
}

%% file: sections/6-case_study.tex
\section{Case Study}

In this section, we conducted two case studies on two popular quantum algorithms, \textit{i.e.}, 
Grover's Algorithm~\cite{grover1996fast} and Quantum Fourier Transform (\textit{QFT}) algorithm~\cite{coppersmith2002approximate},
to demonstrate the usefulness of \toolName.
The users involved in the case studies are two quantum computing experts (E12 and E3) who also participate in the expert interviews in Section \ref{sec:expert_interview}.
Also, all experts were asked to use a monitor with a resolution of $1920 \times 1080$ beforehand.

\newcommand{\rom}[1]{\uppercase\expandafter{\romannumeral #1\relax}}

\subsection{Case Study \rom{1} - Grover's Algorithm}

\definecolor{module_storage}{RGB}{159, 168, 69}
\definecolor{module_processing}{RGB}{159, 183, 172}
\definecolor{module_visualization}{RGB}{178, 78, 50}

\begin{figure*}[t]% specify a combination of t, b, p, or h for top, bottom, on its own page, or here
  \centering % avoid the use of \begin{center}...\end{center} and use \centering instead (more compact)
  \includegraphics[width=0.9\linewidth]{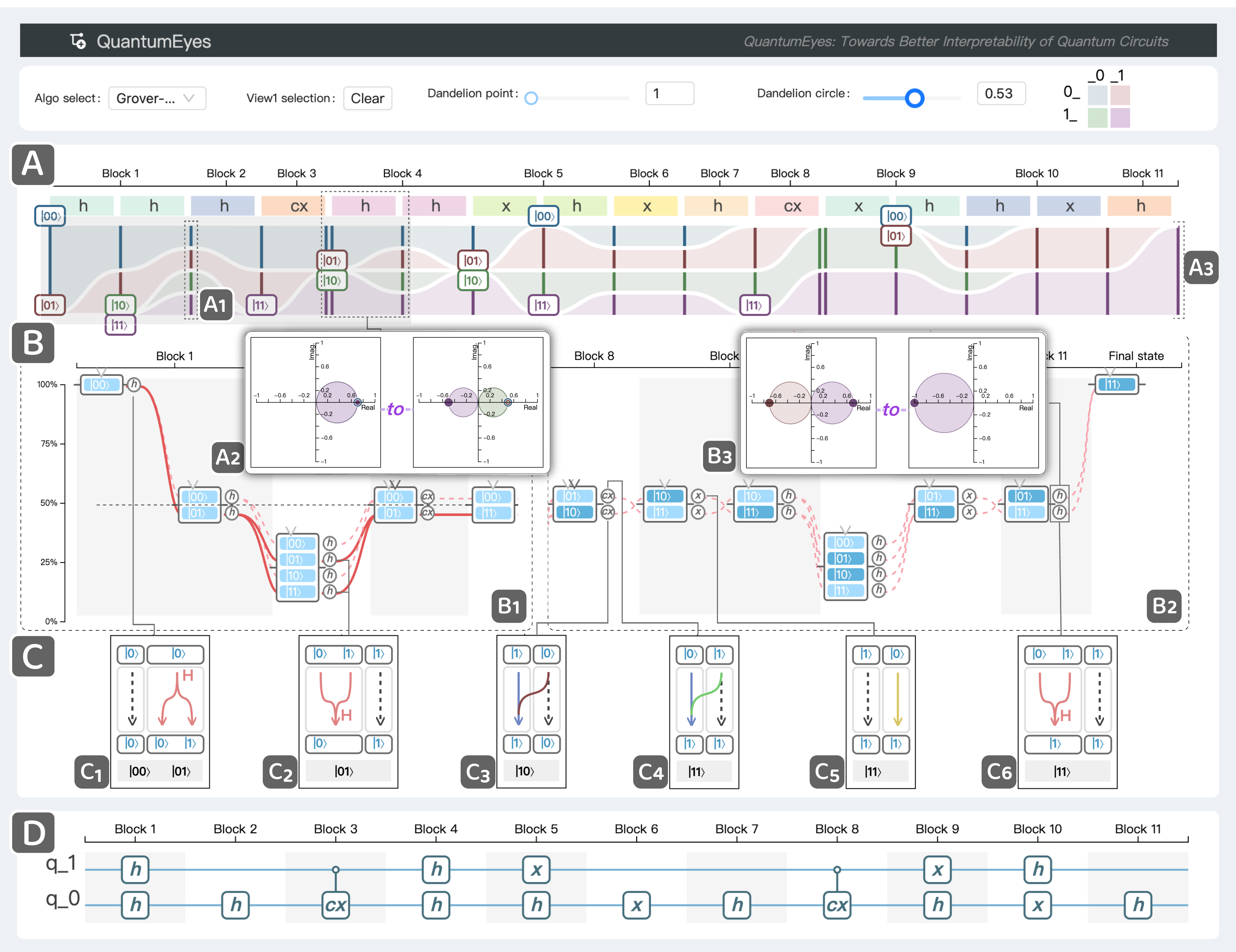}
  \caption{  
  \revise{The visualization system \toolName{} enhances the interpretability of Grover's Algorithm through the global analysis (A-C) to explain the operations of quantum gates and local analysis (A\textsubscript{2} and B\textsubscript{3}) to reveal the implicit reasoning of basis state's measured probability.
% Probability Summary View (A) shows the overview of a quantum circuit regarding the measured probability.
% State Evolution View (B) supports the in-depth analysis of the basis states across each step.
% \revise{Gate Explanation View} (C) explains the effects of quantum gates via the transformation of qubit states.
% State Comparison View (A\textsubscript{2} and B\textsubscript{3}) with the geometrical visualization \designName{} enables users to better understand the measured probability regarding amplitudes.
\revise{The original quantum circuit (D) is given for a better comparison of \toolName{} and quantum circuit diagrams. 
The execution order for the two-gate block is from the qubit with a higher number to a lower number (\textit{e.g.}, from $q_{1}$ to $q_{0}$).
}}}
  \label{fig:case1}
\end{figure*}

\revise{
Grover's algorithm
% ~\cite{toyama2013quantum}
\cite{grover1996fast} 
% \yong{1. Pls add the original citation here. 2. Pls double check the whole sentence.}
is a quantum computing algorithm for searching an unsorted database, which is shown to be more efficient than classical algorithms. It works by repeatedly applying a process called amplitude amplification, which increases the probability of selecting the correct item(s) and decreases the probability of other items.}
We worked with E12, whose research interest includes applying Grover's Algorithm to speed up the unstructured searching problems.
To find more insights behind the quantum circuit used in his research,
%everyday routine tasks, 
E12 leveraged \toolName\ to interactively explore Grover's Algorithm.
Following the example~\cite{li2023qasmbench}, we implemented a 2-qubit Grover's Algorithm for the study.

\textbf{Identifying the functionality block from the visualization. }
E12 began by examining the Probability Summary View and quickly noticed that the probability of State $\ket{00}$ was the largest at the beginning of the circuit. However, this dominance gradually diminished and was replaced by State $\ket{11}$ eventually. 
% This transition can be observed through the length of the line segments and stacked areas at the final step . 
He noted that this transition occurred due to the functionality block of \textit{amplitude amplification} identified by the stacked areas (Fig. \ref{fig:case1}\subcomponent{A\textsubscript{3}}), despite having no prior knowledge of the specific basis state being sought (\textbf{R1}).
Bearing this in mind, E12 became curious about the other functionality blocks of Grover's Algorithm, \textit{i.e.}, the initialization and the \textit{oracle}.
With the clear goal, E12 found that the probabilities of the four basis states were identical, each having a probability of 0.25 as shown in Fig. \ref{fig:case1}\subcomponent{A\textsubscript{1}}. E12 identified the step following the two Hadamard gates (\textit{i.e.}, H gates) at the end of the initialization, as all basis states are in a state of superposition with equal probability, precisely reflecting the characteristic of the initialization.
E12 then noticed that the identified initialization was succeeded by a gate sequence of the ``H-CX-H'' combination.
\textit{``These three gates are commonly employed as an oracle that flips the signs of states, but I still have doubts about this and require further clarification to confirm my understanding.''}
We directed E12's attention to the \designName\ for analyzing the amplitudes (\textbf{R2, R4}).
Utilizing this function, E12 discovered the amplitudes of State $\ket{11}$ were flipped to negative values(Fig. \ref{fig:case1}\subcomponent{A\textsubscript{2}}).
\textit{``This is precisely what I anticipated. The flip of State $\ket{11}$ aligns with the findings of the target state we speculated earlier. Moreover, the flip of the amplitude confirmed that the three quantum gates are an oracle for sure.''}

%%%%%%%%%%%%%%%%%%%%%%%% change the title %%%%%%%%%%%%%%%%%%%
\textbf{Uncovering the facts of the initialization and oracle. }
After the identification, E12 started to perform an in-depth analysis of each functionality block.
By brushing the steps of initialization and the oracle from the stacked area chart,
the State Evolution View was displayed as shown in Fig. \ref{fig:case1}\subcomponent{B\textsubscript{1}}.
To delve further into the quantum gate's effect from a high-level perspective (\textbf{R2}), E12 clicked the ``h'' symbols of the Hadamard gate and displayed the visual explanations of the Hadamard gates (Fig. \ref{fig:case1}\subcomponent{C\textsubscript{1}}).
Taking a close look at the appended view, E12 observed that
% the two qubit states (\textit{i.e.}, $\ket{0}$ and $\ket{0}$) were clearly illustrated by the lines in the second row.
% Specifically, he clarified that 
the first qubit stayed still without any operation, while the second qubit was split into two states $\ket{0}$ and $\ket{1}$ in superposition.
E12 commented \textit{``I am truly impressed that this visualization can easily explain why the final states are $\ket{00}$ and $\ket{01}$ through the decomposition of the gate operation.''}
% I could now understand how Hadamard gate performs in the initialization process.''

The expert then moved on to the analysis of the oracle, which is used to flip the signs of the target state (\textit{i.e.}, $\ket{11}$ in Block 4).
\textit{``I am curious about how the target $\ket{11}$ was generated before the amplitude process''} (\textbf{R3}), E12 commented.
Through hovering over the target state $\ket{11}$, E12 noticed an eye-catching red path to indicate how the $\ket{11}$ was generated, as shown in Fig. \ref{fig:case1}\subcomponent{B\textsubscript{1}}.
He could confidently identify that the States $\ket{01}$ and $\ket{11}$ are the origin of the evolution, which were merged by the following Hadamard gate (Fig. \ref{fig:case1}\subcomponent{C\textsubscript{2}}).
% Then the  state was acted on by a CNOT gate, followed by an H gate to flip the sign of the target state $\ket{11}$.
% \textit{``To my surprise, I can easily identify how the target state $\ket{11}$ is created from the visualizations of two Hadamard gates} (Fig. \ref{fig:case1}\subcomponent{A\textsubscript{2}} and Fig. \ref{fig:case1}\subcomponent{C\textsubscript{2}}).\textit{''}

% He confirmed that the trace-back design is adequate to analyze how a specific state evolved, 

\textbf{Exploring the hidden insights of the amplification. }
E12 proceeded to analyze the functionality block of amplification, which is employed to amplify the probability of the flipped target state.
By brushing the corresponding steps in the stacked area chart,
% E12 thoroughly examined each step to understand how the amplitude works.
E12 got a quick intuition of the operations of the CNOT gate (Fig. \ref{fig:case1}\subcomponent{C\textsubscript{3}} and \subcomponent{C\textsubscript{4}}) and the NOT gate (Fig. \ref{fig:case1}\subcomponent{C\textsubscript{5}}).
% commented, \textit{``the designs provide me with crucial information about the quantum circuit''}.
% For example, the effect of the CNOT gate (Fig. \ref{fig:case1}\subcomponent{C\textsubscript{3}} and \subcomponent{C\textsubscript{4}}) and the NOT gate (Fig. \ref{fig:case1}\subcomponent{C\textsubscript{5}}).
To determine the reason for the sudden increase in the probability of State $\ket{11}$ (\textbf{R2}), E12 took a glance at Fig. \ref{fig:case1}\subcomponent{C\textsubscript{6}} and quickly noticed that the Hadamard gate merged the first qubit's state (\textit{i.e.}, $\ket{0}$ and $\ket{1}$) and generated a new qubit state (\textit{i.e.}, $\ket{1}$).
\textit{``This is mainly because the first state $\ket{0}$ is negative, leading to the new state of  $\ket{1}$ other than  $\ket{0}$''}, E12 said, \textit{``However, I cannot still understand why the probability changes into 1 instead of other numbers''} (\textbf{R4}).
% I hope the visualization provides a `quantitative' explanation of the number 1.''

Thus, as hinted by us, E12 further moved to the \designName\ of the step by clicking the last step's background.
After a glance, he noticed there are two states (\textit{i.e.}, $\ket{01}$ and $\ket{11}$) at the left system and only one state (\textit{i.e.}, $\ket{11}$) at the right with a symbol denoting the operation gate (Fig. \ref{fig:case1}\subcomponent{B\textsubscript{3}}).
% It is apparent that the State $\ket{01}$ disappears through the gate transformation.
E12 found that the imaginary parts of all states are zero, as indicated by their zero y-coordinates. 
Furthermore, the real part of State $\ket{11}$'s amplitudes changed from around 0.7 in the left chart to -1.0 in the right-hand chart.
% , leading to the sudden increase of the circle's area.
\textit{``I am surprised that the \designName\ tells me that the flip of phase did not cause the change of the probability, the real part of the amplitude actually makes the state's probability two times its initial state.''}

\subsection{Case Study \rom{2} - Quantum Fourier Transform Algorithm}

\begin{figure*}[t]% specify a combination of t, b, p, or h for top, bottom, on its own page, or here
  \centering % avoid the use of \begin{center}...\end{center} and use \centering instead (more compact)
  \includegraphics[width=0.9\linewidth]{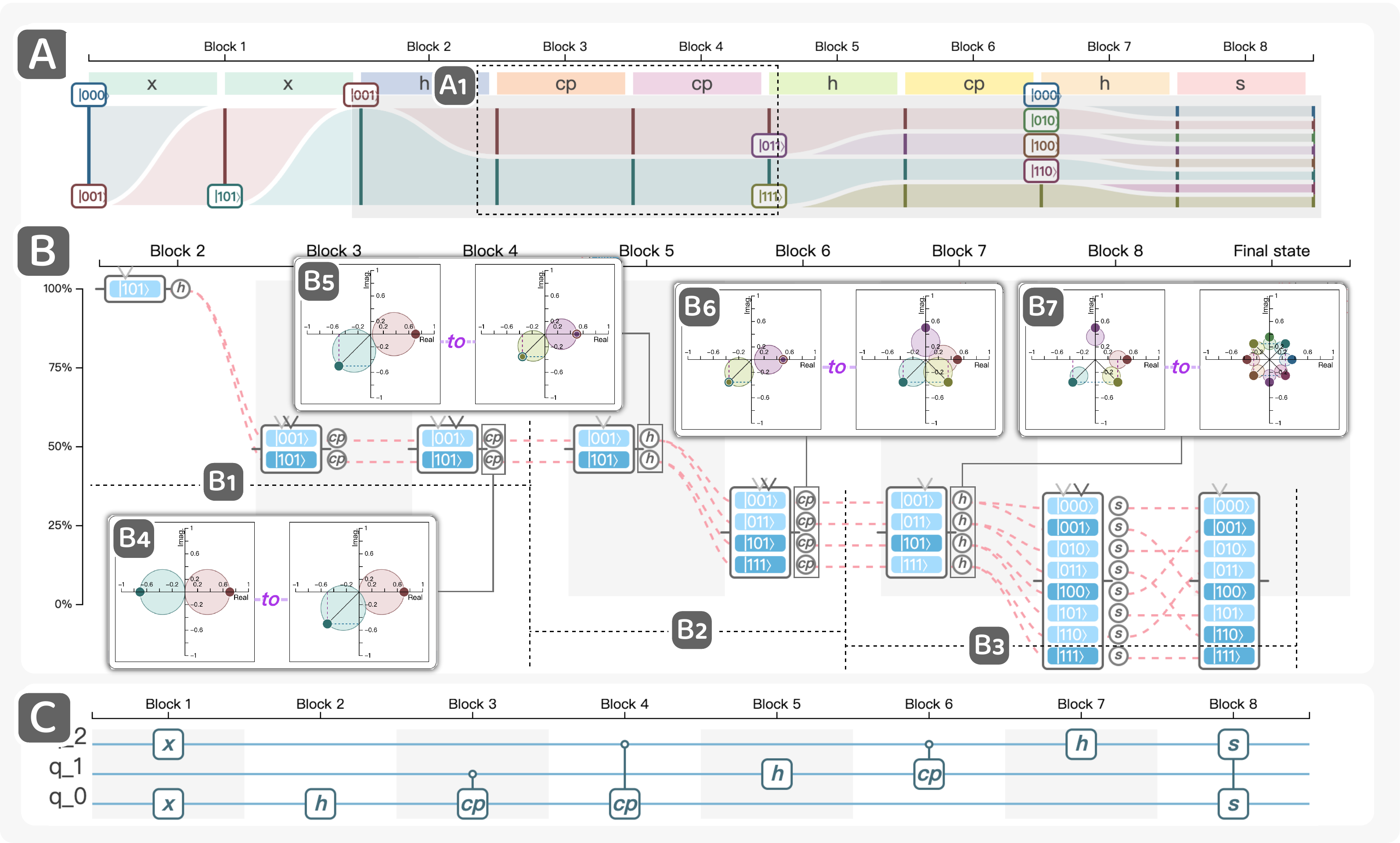}
  \caption{
  \revise{
    The case for Quantum Fourier Transform algorithm.
  Three coordinated views (A-C) visualize the development of basis states for the global analysis,
  while  \designName\ (B\textsubscript{4}-B\textsubscript{7}) explains how measured probabilities are determined by amplitudes for the local analysis.
  The \textit{CP} symbols in the quantum circuit diagram (C) indicate the Controlled-Phase gates and \textit{S} symbols indicate the SWAP gates.
  }
  % (A) Probability Summary View depicts the architecture of the algorithm using the measured probability of each basis state.
  % (B) State Evolution View 
  }
  \label{fig:case2}
\end{figure*}

We worked with E3, whose research direction is Quantum Uncertainty, to understand a widely-used quantum algorithm, \textit{i.e.}, Quantum Fourier Transform (\textit{a.k.a.}, QFT) algorithm~\cite{shor1994algorithms}. 
% \yong{pls add a citation here.}
%
\revise{The QFT algorithm converts the amplitudes of a quantum state into the 
corresponding values in the
frequency domain, which is similar to what the classical Fourier Transform does with signals. It forms a foundation for other quantum algorithms, such as Shor's Algorithm~\cite{shor1999polynomial}.}
% QFT is a quantum-version of a Fourier transform which maps a quantum state to a linear combination $\sum_{i=0}^{N-1}y_{i}\ket{x}$.
% \revise{The QFT algorithm is a linear transformation that operates on quantum bits, mapping quantum states in a manner similar to how the Fourier Transform acts on classical bits}
We implemented the quantum circuit following the guidelines of \textit{Qiskit}~\cite{qiskit4qft}.
% and presented \toolName\ to E3 to perform our study.

\textbf{Understanding the architecture of QFT algorithm. }
The expert E3 started by brushing the whole quantum circuit from the probability overview because he thought the QFT algorithm is an entity that cannot be split into different functionality blocks.
Indicated by the first two X gates with probabilities of 1.0, E3 commented, \textit{``These two X gates are for the state preparation because the lengths of the line segments remain the same during Block 1.'' }
Meanwhile, he speculated the number to be mapped is 5 due to the decimal of State $\ket{101}$.
After identifying the number to be mapped, E3 started to investigate the quantum circuit architecture of the QFT algorithm (\textbf{R2}).
By exploring the State Evolution View along with the original circuits, E3 quickly found that the three key processes of the algorithm 
\textit{``I can easily identify the three iterations of QFT }(as shown in Fig. \ref{fig:case2}\subcomponent{B\textsubscript{1}}, Fig. \ref{fig:case2}\subcomponent{B\textsubscript{2}}, and Fig. \ref{fig:case2}\subcomponent{B\textsubscript{3}}) \textit{from the three continuous processes with the downward trend of probabilities from the middle view,}
% decreases of the states' probabilities shows the three combinations of the quantum circuit (\textit{i.e.}, H gate and Controlled Phase gate), 
\textit{each making the probability drop to 0.25''.}
He also praised the advantage of the evolution view to intuitively reveal the temporal change of states' probabilities along the circuit, making the analysis of the gate's functionality more efficient and smooth.

\textbf{Disclosing the implicit reasons of the measured probability. }
E3 then glanced at the probability summary of the QFT algorithm and found that the probability of the two States $\ket{001}$ and $\ket{101}$ did not change after the two Controlled-Phase gates (Fig. \ref{fig:case2}\subcomponent{A\textsubscript{1}}). Thus, he planned to find more hidden insights about this phenomenon by drilling down to the local analysis using \designName (\textbf{R4, R5}).
According to the geometrical representation of Fig. \ref{fig:case1}\subcomponent{B\textsubscript{4}}, E3 noticed that the circle of State $\ket{101}$ rotates around 45 degrees anticlockwise after the Controlled-Phase gate, making the amplitudes change but preserving the circle area.
\textit{``This design is fascinating to me because I can analyze the gate's effect from a perspective of geometry intuitively.''}
Next, he clicked the following Hadamard gate to find the reason for superposition using the \designName.
From Fig. \ref{fig:case2}\subcomponent{B\textsubscript{5}}, E3 noticed the both of the two original States $\ket{101}$ and $\ket{001}$ became two smaller circles.
\textit{``Before today, I can only observe the four states with the same probability of 0.25 after Hadamard gates. It is brilliant to build a mathematical intuition of the measured probability and the amplitudes.''}
After analyzing the individual quantum gate, E3 planned to investigate how the QFT algorithm represents a random quantum state by a series of continuous basis states (\textit{i.e.}, $\ket{000} \cdots \ket{111}$).
Hence, E3 clicked the last two quantum gates before the final SWAP gate and then adjusted the radius to separate all circles (\textbf{R6}), as shown in Fig. \ref{fig:case2}\subcomponent{B\textsubscript{6}} and Fig. \ref{fig:case2}\subcomponent{B\textsubscript{7}}.
\textit{``This actually matches what I expected,''} E3 commented
\textit{``From the first chart, I realized that the Controlled-Phase gate can only `rotate' a state but never `separate' a state into multiple states.''}
E3 further noticed that the four states are located in four different directions (\textit{i.e.}, cardinal directions and diagonal) (Fig. \ref{fig:case2}\subcomponent{B\textsubscript{6}}).
And then, the Hadamard gate generates each state into a new basis state in the opposite direction (Fig. \ref{fig:case2}\subcomponent{B\textsubscript{7}}), making it possible to handle eight basis states with smaller circle area.
% (\textit{i.e.}, $N = 2^{3}$).
\textit{``The \designName\ provides me a holistic picture of how the quantum gate changes the amplitudes of basis states, which makes the analysis of amplitudes more effective than ever before.''}

%% file: sections/7-expert_interview.tex
\section{Expert Interview}
\label{sec:expert_interview}

We further conduct a well-designed interview with actual domain experts to demonstrate the effectiveness and usability of \toolName\ and the embedded \designName.

\subsection{Study Design}

\textbf{\revise{Participants and apparatus.} }
We recruited 12 domain experts (\textbf{E1-12}) (12 males, $age_{mean}=34.0$, $age_{sd}=5.8$) from 6 different educational institutions (\textbf{E1-12}) in the U.S. to join our in-depth expert interview.
% \revise{We selected the gender of participants without any preference.}
% None of the above experts participated in the study for collecting design requirements.
\revise{
These participants were selected by mainly considering their research background in quantum computing and checking whether they have relevant research experience, guaranteeing the reliability of the collected feedback. 
% We selected the participants
% \st{based on their 
% accessibility and }
% % qualifications relevant to the topic of the study.
% by considering their
% familiarity with quantum computing.
}
% \yong{What is their accessibility?}
%
%
More specifically, five participants (\textbf{E1, E9-12}) are working on Quantum Error Mitigation, six experts (\textbf{E4-7, E8, E13-14}) study Quantum Machine Learning (QML), two experts (\textbf{E2-3}) are working on Quantum Uncertainty, and one expert (\textbf{E7})'s research direction is Quantum System Design.
All participants have an average of 5.9 years of experience in quantum computing.
% To guarantee the generalizability of the study, none of the experts has a background in data visualization and HCI.
The interview was conducted via the online Zoom meeting with a $1920 \times 1080$ resolution monitor.

% \textbf{Presetting. }
% We provide a commonly-used quantum algorithm, \textit{i.e.}, a 2-qubit Grover's Algorithm~\cite{2qgrover}, to support the exploration  of \toolName.
% Also, we prepared six carefully-designed tasks regarding the effectiveness evaluation of all visual designs (see \ref{sec:tasks}).

% \begin{table}[t]
% \centering
% % \begin{tabular}{c|p{0.8\columnwidth}}
% \begin{tabular}{c|p{0.8\columnwidth}}
% \hline
% T1 & Identify the overall trends of all basis states regarding the measured probability. \\ \cline{2-2} 
% T2 & Identify the generation of the basis states via the trace-back analysis.        \\ \cline{2-2} 
% T3 & Describe the operations and effects of the quantum gates by comparing the basis states before and after.    \\ \cline{2-2} 
% T4 & Explain the effect of the quantum gates from the perspective of the qubit state analysis.            \\ \hline
% T5 & Explain how quantum gates change the amplitudes of the basis states.       \\ \cline{2-2} 
% T6 & Explain how amplitudes of basis states change the corresponding measured probabilities.             \\ \hline
% \end{tabular}
% \caption{%
% All pre-defined tasks are grouped into two categories, \textit{i.e.}, the effectiveness evaluation of visual designs for global analysis (\textbf{T1-4}) and local analysis (\textbf{T5-6}).
% }
% \label{table:1}
% \end{table}

\textbf{Procedures.}
The study was conducted on the online system \toolName.
We carried out the one-on-one, semi-structured study for all experts.
Specifically, we first introduced the visual design of all views of \toolName\ along with \designName.
Afterward, 
we invited all participants to accomplish six pre-defined tasks using \toolName{}.
\revise{The first four tasks are designed to evaluate the effectiveness of \toolName{} for global analysis, including analyzing the overall trend of basis state probabilities, identifying the gate effect, and explaining the gate effect in terms of the changes of basis states and qubit states.
The remaining two tasks aim to evaluate the effectiveness of the dandelion chart's effectiveness for local analysis.}
The detailed task list can be found in Appendix \ref{sec:tasks}.

% (see \ref{sec:tasks})
% % \st{of exploring the examples (\textit{i.e.}, \textit{Grover\_n2\_Qiskit})}
% using \toolName{}, where the tasks are labeled as T1-T6.
% \yong{Need a brief introduction to the questions.}
% \yong{why are ``the examples'' consisting of only one item here?}
% \revise{
% Specifically, T1-4 are designed to evaluate the purpose of each view for global analysis (i.e., T1: overall trend perception, T2-3: gate effect explanation, T4: fine-grained gate explanation regarding individual qubit), and T5-6 are proposed to test dandelion chart's effectiveness of measured probability explanation. 
% \yong{I would suggest that we should either add more details here or add the exact table to the paper instead of putting it in the appendix.}
%
%
We then asked them to verbally explain how the quantum states are evolving across the quantum circuit.
% }
% Additionally, they were asked to describe how quantum states are modified by the various quantum gates concerning the amplitudes and probability for the local analysis. 
% selected in the previous circuit and explain how the amplitudes determine the probability.
The aforementioned process lasts approximately 40 minutes.
After completing the tasks, all participants were encouraged to provide feedback on all the proposed visual designs in a think-aloud manner.
Furthermore, followed by prior work~\cite{ruan2023isualization}, \revise{we also invited participants to rate \toolName{} using a 7-point Likert scale based on the post-study questionnaire (Table~\ref{table:2}) regarding each aspect of design requirements we collected beforehand.}
% based on questionnaires shown in \ref{table:2},
% which can effectively assess the quantum computing-specific visualization system in prior work~\cite{ruan2023isualization}.} \yong{What??? Why assess the visualization in prior work??? }
The post-study interview lasted approximately 20 minutes, during which we recorded and took notes about the entire study process.

\begin{table}[t]
\caption{%
The questionnaire consists of four parts, \textit{i.e.}, the effectiveness (\textbf{Q1-4}), usability (\textbf{Q5-7}),  user interaction (\textbf{Q8-9}), and  visual designs (\textbf{Q10-12}).
}
\centering
% \begin{tabular}{c|p{0.8\columnwidth}}
\begin{tabular}{c|p{0.8\columnwidth}}
\hline
\multicolumn{1}{c|}{Q1} & The workflow of global and local analysis can explain the quantum circuits comprehensively.                                                                                      \\ \cline{2-2} 
\multicolumn{1}{c|}{Q2} & The system can effectively support the evolution analysis of each basis state.                                                                                                   \\ \cline{2-2} 
\multicolumn{1}{c|}{Q3} & The system can intuitively explain the gate effect via the visualization of qubit states.                                                                                         \\ \cline{2-2} 
\multicolumn{1}{c|}{Q4} & The dandelion chart can effectively explain the measured probability based on the amplitudes.                                              \\ \hline
\multicolumn{1}{c|}{Q5} & The system is easy to learn.                                                                                                                                                      \\ \cline{2-2} 
\multicolumn{1}{c|}{Q6} & The publicly-available QuantumEyes system is helpful for domain users.                                                                                                           \\ \cline{2-2} 
Q7                      & I would like to use the QuantumEyes system to better understand quantum circuits in the future.                                                                                   \\ \hline
Q8                      & The user interaction of the system is smooth.                                                                                                                                    \\ \cline{2-2} 
Q9                      & The user interaction is easy to use for domain users.                                                                                                                            \\ \hline
Q10                     & The overall design is easy to understand.                                                                                                                                         \\ \cline{2-2} 
Q11                     & For global analysis, the three coordinated views are helpful in understanding the effects of quantum gates. \\ \cline{2-2} 
Q12                     & For local analysis, the dandelion chart is useful to visualize how amplitudes affect the probability intrinsically.                                                              \\ \hline
\end{tabular}
\label{table:2}
\end{table}

\begin{figure}[t]% specify a combination of t, b, p, or h for top, bottom, on its own page, or here
  \centering % avoid the use of \begin{center}...\end{center} and use \centering instead (more compact)
  \includegraphics[width=\columnwidth]{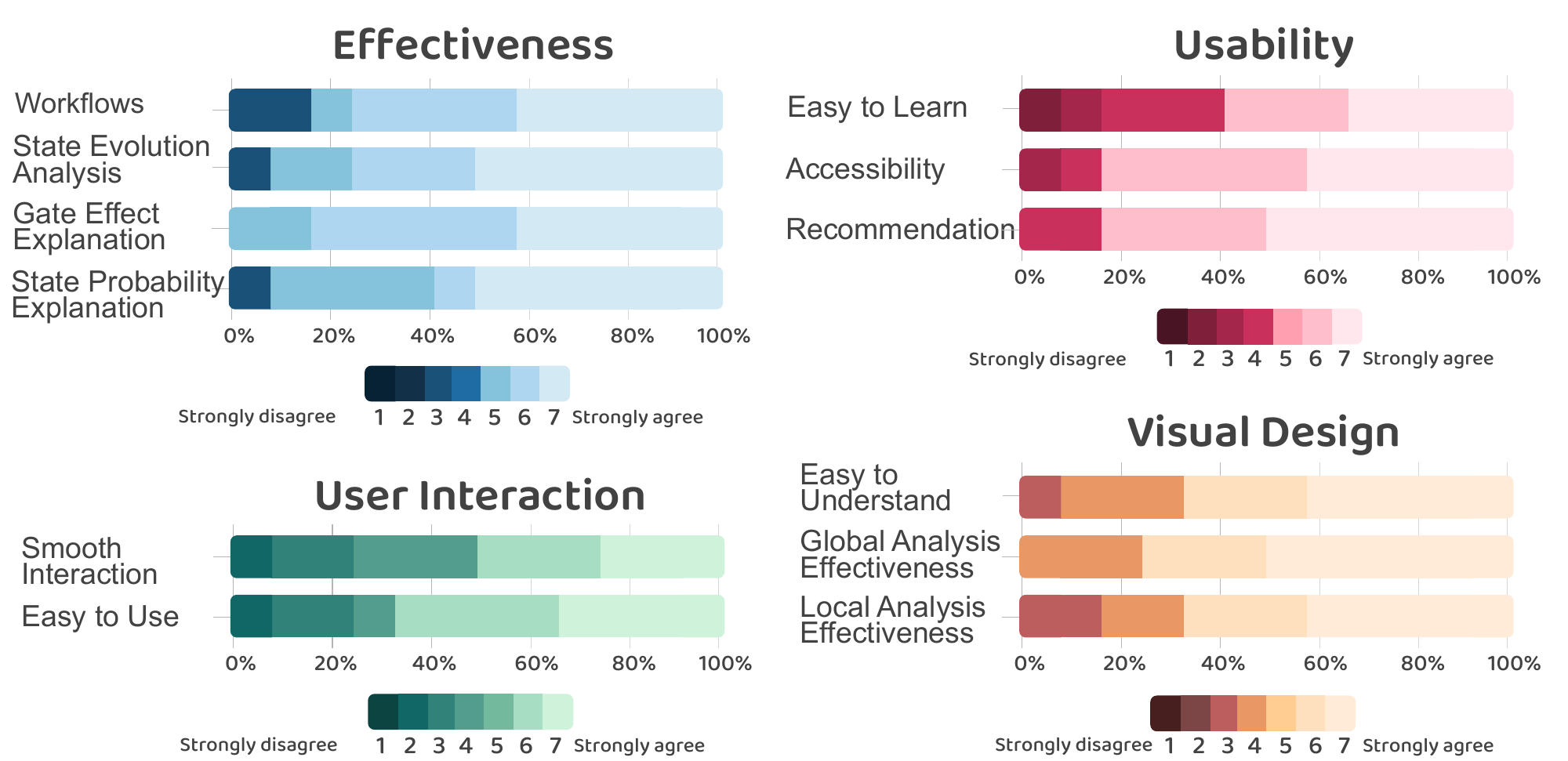}
  \caption{\revise{The summary of the feedback of the questionnaire.}}
  \label{fig:6}
\end{figure}

\subsection{Result}

We summarized all collected feedback regarding the four aforementioned aspects of the evaluation
as Fig. \ref{fig:6}.

\textbf{Effectiveness. }
Most participants appreciated the effectiveness of \toolName\ to enhance the interpretability of quantum circuits ($rating_{mean}=6.02, rating_{sd}=1.18$).
\textbf{E3-6} agreed that the workflow of global and local analysis is exactly what quantum computing users expect to see to explain the effects of quantum gates.
% \textbf{E5} commented, \textit{``The global analysis provides me with a high-level summarization of gate effect, at the same time, the local analysis can give users an in-depth explanation of the quantum states. I believe the two aspects can make the users understand a specific circuit comprehensively.''}
Meanwhile, the trace-back analysis is praised by \textbf{E1-2} and \textbf{E9}.
\textit{``I need to manually calculate the state vectors to figure out how a state was produced and developed in my daily work before. 
This function provided by \toolName\ is really fascinating to me''}, E2 said.
% Also, \textbf{E7-8} expected to use \toolName\ for educational purposes because the explanation of the qubit state can intuitively tell how the quantum gate operates a quantum state.
Furthermore, most participants (\textbf{E1-7, E9-11}) highly appreciate the novel design \designName, \textit{``which is helpful to grasp the measured probability of basis states.''}
\textbf{E11} also commented that \toolName\ can help him with circuit design and debugging due to the intuitive visualization of probability regarding the state's amplitudes.
% provides an . This property.

\textbf{Usability. }
The majority of participants applauded the usability of \toolName\ in interactively enhancing the quantum circuit's interpretability ($rating_{mean}=5.88, rating_{sd}=1.65$).
\textbf{E2-5} mentioned that the visualization system is user-friendly for quantum computing researchers and learners.
Among them, \textbf{E4} commented, \textit{``I can easily interact with the interface and accomplish all tasks, even though I do not have any background in visualization before.''}
\textbf{E7} and \textbf{E12} emphasized that they prefer the easy-to-understand visualizations, and \toolName\ indeed provides the visualizations that they are familiar with in their daily routine tasks, like the original quantum circuit diagram and the visualization of the transformation of the qubit states.
% Meanwhile, \textbf{E9-12} 
% highly praised the online system \toolName, which every quantum computing user can access and benefit from.
% Additionally, they also
% strongly recommend we publish the scalable visual design \designName\ as an online package to enable an instant build for quantum computing developers.

\textbf{User interactions. }
Most participants generally agreed that the user interactions in \toolName\ are easy-to-use for quantum computing users ($rating_{mean}=5.54, rating_{sd}=1.66$).
Among all feedback, \textbf{E3-4} and \textbf{E9} gave highly positive feedback for the interactions of decreasing the circle radius in \designName\ to reduce the visual clutter.
\textbf{E3} pointed out that he feels struggle to adopt Bloch Sphere to inspect only the single-qubit state. \revise{Dandelion chart} addresses the limitations perfectly while preserving the characteristic of displaying the quantum amplitudes.
Meanwhile, E9 confirmed that reducing the circle area is feasible \textit{``because users always need to focus on the circle with the largest area.''}
% Meanwhile, \textbf{E9} said, \textit{``the interaction of shrinking the circle can definitely support the analysis of a large number of basis states. Moreover, reducing the circle area with a smaller probability is feasible because I always need to focus on the circle with the largest area.''}
Furthermore, \textbf{E12} also expressed the desire to recommend \toolName\ to his research group members due to the easy-to-use system interactions.
% commented that the user interactions of \toolName\ is smooth and easy to use. He 

\textbf{Visual designs. }
Most participants gave positive feedback about the visual designs in \toolName\ ($rating_{mean}=6.05, rating_{sd}=0.99$).
Specifically, \textbf{E5} mentioned that the visual designs for global analysis are informative.
\textit{``I like the visualization to show the how state evolves because it can directly tell when and how a basis state is developed. This characteristic would truly aid the analysis of  quantum algorithms in my daily tasks.''}
Also, \textbf{E7} was willing to adopt \designName\ for his own local analysis of the quantum states,
% \textit{``To my surprise, this visualization can directly show how the quantum amplitudes determine the probability based on the geometry''}, \textbf{E7} commented, 
\textit{``To my surprise, this design is brilliant because everyone can find the rationale of probability changes without the complex matrix calculation, even for the beginners in quantum computing.''}

\textbf{Suggestions. }
In addition to the positive feedback, several participants also offered constructive suggestions.
% to further improve \toolName.
\textbf{E4} suggested that incorporating a feature to fold and unfold the blocks would be helpful for comparative analysis.
\textbf{E9} also noted that \toolName\ could be extended to visualize the temporal change of parameters in variational quantum circuits.
\textbf{E10} expressed that a transition might be useful to highlight the difference when comparing a pair of \designName \textit{s}.

%% file: sections/8-discussion.tex
\section{Discussion}

In this section, we first summarize the lessons learned during the development of \toolName\ and \designName. Then, we discuss the limitations of our proposed visual designs.

\subsection{Lessons Learned}
We reported the learned lessons from the development of \toolName.

% vis' value in qc（visually correlation of danelion chart)
\textbf{Indispensable necessity of visualization to interpret quantum computing. }
During the process of working with domain experts in the requirement collection and evaluation, 
they confirmed the great importance of interpreting quantum computing using visualization approaches.
According to the actual use of \designName, experts appreciated the impressive design, while also
% which could directly see how amplitudes affect the probability through the correlations of the point's position and the circle's area.
% Specifically, they
pointed out that quantum computing is not transparent for users to learn, which is exactly the domain where visualization can aid.
Thus, they also mentioned that they preferred the designs with linked visual channels to offer the explanation intuitively.
% , other than those simple visualizations with multiple individual elements (e.g., a set of bar charts).

% accessibility for domain users(probability summary, original circuit, intuitive view 3
\textbf{Design considerations tailored for quantum computing users. }
By working with domain experts, we realize that lowering the learning costs of the proposed visual design for domain users is significant.
In our study of design requirements, all participants preferred solutions that were easy to learn, simplifying the paradigm shifts from reading to understanding.
For example, the overview of the probability is appreciated and used as the starting point of the system.
% as they commonly leverage probability for quantum circuit analysis in their tasks.
Also, they praised the implementation of the original quantum circuit in \toolName\ because the comparative analysis with the original circuit can significantly shorten their learning curves.
% , making the interaction with the system more efficient.

\subsection{Limitations \revise{and Future Work}}
\revise{There are still several limitations of \toolName.}
% Our evaluation has shown that \toolName\ and the embedded \designName\ can effectively improve the interpretability of quantum circuits.
% However, it is not without limitations.

% vqc analysis
% \textbf{Interpretability of variational quantum circuits.}
% Our approaches can effectively facilitate domain users' understanding of quantum circuits.

\textbf{Application scope.}
\revise{All the participants highly appreciate the effectiveness of \toolName{} in helping quantum computing developers and users understand the working mechanism of static quantum circuits, which is the widely-used quantum circuits.
%--- The above sentence feels strange. 
% \st{, which focuses on the static quantum circuit, which is the most widely-used type of quantum circuit in quantum computing.}
} 
% However, another type of quantum circuit, i.e., variational quantum circuit (VQC),
% also gains more and more attention. QuantumEyes cannot be directly
% applied to VQC for}, 
% \toolName{} provides quantum computing users with an intuitive and convenient way to understand and explore the working mechanism of static quantum circuits.
However, with the growth of another type of quantum circuits, \textit{i.e.}, variational quantum circuits (VQC), also gain more and more attention.
Although our novel design \designName{} can be seamlessly applied to VQC applications to analyze quantum state evolution, the visual analytics system (\textit{i.e.}, \toolName{}) as a whole cannot be directly applied to VQC for now.
% \toolName{} cannot be directly applied to VQC for now; our novel design \designName{}, however, can be seamlessly applied to the VQC application to support the visual analysis of how quantum states are modified across different iterations.
\revise{In future work, we plan to extend \toolName{} to analyze variational quantum circuits and other features such as the generalization of the visual feature ``Path of state evolution'' of \toolName{} to a ``Path of Bloch coefficient/Pauli probability evolution''. }
% \st{further develop global visualizations to support visual analytics of variation quantum circuits.}
% \yong{Shaolun, pls fill this part.}

\revise{\textbf{Scalability.}}
\revise{
The evaluation has demonstrated that our visualization works well for visualizing the states of two and three qubits.
% states' visualization. 
However, due to the limited screen space, the visualization components in \toolName{} for global analysis, \textit{i.e.}, the Probability Summary View, State Evolution View and Gate Explanation View, may suffer from scalability issues with the increase of qubits in quantum circuits.
% The components in \toolName{} for global analysis may suffer from scalability issues when the number of qubits exceeds 5.
% \st{However, the} \designName{} \st{can perform better than the other views by scaling down the radii of circles and can show a group of circles with a larger circle area.}
The visualization component of \toolName{} for local analysis, \textit{i.e.}, \designName{}, has better scalability than the above three views, as it can reduce the radii of circles to explain the measured probabilities of more basis states. But when there are a large number of qubits in the quantum circuits, visual clutters may also appear.
In the future, it is worth further exploration on how to enhance the scalability of \toolName{}.}

% noise analysis
% \textbf{Noise analysis in real quantum computers.}
% Our proposed visualization approaches can improve the interpretability of quantum circuits.
% However, we did not consider noise in quantum computers as the system is based on quantum simulators.
% Thus, the object of the study is noise-free quantum circuits, which are more appropriate for interpretability enhancement without the influence of the noise hidden in real quantum computers.
% \revise{We aim to incorporate noise models into the system to simulate real quantum
% computers with the influence of noises in the future.}

\textbf{\revise{User-friendly interactions.}}
\revise{The availability of \toolName{} gained positive feedback from all participants. 
However, there are still several limitations regarding the user interactions we collected during the interview.
First, to aid the obscure connection between the State Evolution View and the original circuit, we plan to implement the folding and unfolding of basis states and their corresponding quantum gates.
% \yong{1. Why this point? 2. What are the limitations here? Pls discuss it first before you talk about the future work!}
% In the future, we aim to improve the usability of \toolName{} to enable more user-friendly interactions. 
% \st{Thus, we plan to implement the folding feature to facilitate comparative analysis and also enhance scalability.} \yong{Why? It seems to come from nowhere.}
%
%
Also, the comparison of two quantum states in \designName{} can be improved by utilizing the transition of circles to highlight the effect of quantum gates.
% Furthermore, \toolName\ is expected to enable the functionality of simulator customization and circuit data loading to enhance the system's flexibility.
Further, to address the issue that the entangled states sometimes cannot be clearly distinguishable in the \designName, we plan to add extra visual elements to highlight those basis states that are entangled using sector length distributions.
Last, \toolName\ is expected to enable the functionality of simulator customization and circuit data loading to enhance the system's flexibility.
}

% * You are also sacrificing the ability to visualize entanglement, e.g., the fully separable state $\vert -,- \rangle = (1,-1,-1,1)/2$ is not clearly distinguishable from the maximally entangled state $CZ \vert +,+ \rangle = (1,1,1,-1)/2$. Note that sector length distributions are a useful tool to visualize such differences.
% Finally, note that the visual feature "Path of state evolution" of QuantumEyes nicely generalizes to a "Path of Bloch coefficient/Pauli probability evolution":
% For every $n$-qubit gate $U$, the coefficient $r_P$ distributes into other coefficients $r'_{P'}$ via $r_P UPU^\dagger = \sum_{P'} r'_{P'} P'$.
% Note that Clifford gates $U$ only permute the Pauli operators, \textit{i.e.}, the coefficients are not split but only reshuffled here.
% For a non-Clifford gate $V$, on the other hand, there exists at least one Pauli operator $P$ for which $VPV^{\dagger}$ is not a Pauli operator, by definition of the Clifford group.
% Please consider incorporating such features in your future work. (This is certainly worthwhile but well beyond the scope of the present submission.)

%% file: sections/9-conclusion.tex
\section{Conclusion}

We present \toolName\, an interactive visualization system to enhance the interpretability of quantum circuits.
By working closely with domain experts, we formulate six design requirements in terms of two analysis levels to guide the design of our system.
Specifically, we propose three coordinated views (\textit{i.e.}, 
a Probability Summary View, a State Evolution View, and a \revise{Gate Explanation View}) to support the global analysis of the quantum state evolution over the whole quantum circuit.
Further, we \revise{propose} a novel geometrical visual design \designName\ for local analysis, enabling users to visually analyze the correlation of basis states' probability and amplitudes based on geometry principles.
We \revise{conduct} two case studies and expert interviews to demonstrate the effectiveness and usability of the proposed visualization approaches.
The result shows that our approaches can effectively facilitate domain users to better understand quantum circuits.

% \toolName\ along with \designName\ to support the interpretability enhancement of quantum circuits.
% To inform the designs, we identified six design requirements together with six quantum computing experts.
% We developed a visualization system \toolName\ to 
% support the global analysis of the quantum state evolution over the whole quantum circuit.
% Also, we proposed a geometrical visual design \designName\ for local analysis, enabling users to visually analyze the correlation of basis states' probability and amplitudes based on geometry principles.
% We conducted two case studies and expert interviews to demonstrate the effectiveness and usability of the proposed visualization approaches.
% The result shows that our approaches can effectively facilitate domain users to better understand quantum circuits.

% In future work, we will explore how \toolName\ can be extended to support visual analytics of variation quantum circuits, satisfying the increasing demand of Quantum Machine Learning.
% Also, we plan to incorporate noise models into the system to simulate the real quantum computers with the influence of noises. 